%% file: main.tex
\documentclass{article}

\usepackage{arxiv}

\usepackage[utf8]{inputenc} 
\usepackage[T1]{fontenc}    
\usepackage{hyperref}       
\usepackage{url}            
\usepackage{booktabs}       
\usepackage{amsfonts}       
\usepackage{nicefrac}       
\usepackage{microtype}      
\usepackage{lipsum}		
\usepackage{graphicx}
\usepackage{natbib}
\usepackage{doi}

\usepackage{amsmath}
\usepackage{multirow}
\usepackage{xurl}
\usepackage[english]{babel}
\usepackage[autostyle, english = american]{csquotes}
\MakeOuterQuote{"}
\usepackage[table,svgnames,dvipsnames]{xcolor}
\newcommand{\highlighttextbf}[1]{{#1}}
\newcommand{\highlightcell}[1]{\cellcolor{LightGreen!50}\highlighttextbf{#1}}

\newcommand\shl[1]{%
  \bgroup
  \hskip0pt\color{cyan!80!green}%
  #1%
  \egroup
}

\usepackage{bibunits} 

\setcitestyle{super} 

\input{macros}

\title{Resolution deficits drive simulator sickness and compromise reading performance in virtual environments}

\author{
	\href{https://orcid.org/0000-0002-1990-1293}{\includegraphics[scale=0.06]{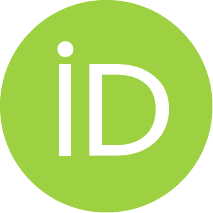}\hspace{1mm}Jialin~Wang}\\
	Computational Media and Arts Thrust\\
	Hong Kong University of Science and Technology (Guangzhou) (HKUST-GZ)\\
	Guangzhou, China\\
	\And
	\href{https://orcid.org/0009-0005-5084-6391}{\includegraphics[scale=0.06]{orcid.pdf}\hspace{1mm}Xinru~Cheng}\\
	Computational Media and Arts Thrust\\
	HKUST-GZ\\
	Guangzhou, China\\
	\And
	\href{https://orcid.org/0009-0009-7913-5437}{\includegraphics[scale=0.06]{orcid.pdf}\hspace{1mm}Boyong~Hou}\\
    AI Thrust\\
	HKUST-GZ\\
	Guangzhou, China\\
	\And
	\href{https://orcid.org/0000-0003-3600-8955}{\includegraphics[scale=0.06]{orcid.pdf}\hspace{1mm}Hai-Ning~Liang}\\
	Computational Media and Arts Thrust\\
	HKUST-GZ\\
	Guangzhou, China\\
	\texttt{hainingliang@hkust-gz.edu.cn}\\
}

\renewcommand{\shorttitle}{Resolution, Sickness, and Reading in XR}

\hypersetup{
pdftitle={Resolution deficits drive simulator sickness and compromise reading performance in virtual environments},
pdfauthor={Jialin~Wang, Xinru~Cheng, Boyong~Hou, Hai-Ning~Liang},
pdfkeywords={text legibility, visual acuity, simulator sickness, extended reality, head-mounted display, effective resolution, reading performance},
}

\begin{document}
\maketitle

\begin{abstract}
Extended reality (XR) is evolving into a general-purpose computing platform, yet its adoption for productivity is hindered by visual fatigue and simulator sickness. While these symptoms are often attributed to latency or motion conflicts, the precise impact of textual clarity on physiological comfort remains undefined. Here we show that sub-optimal effective resolution, the clarity that reaches the eye after the full display-optics-rendering pipeline, is a primary driver of simulator sickness during reading tasks in both virtual reality and video see-through environments. By systematically manipulating end-to-end effective resolution on a unified logMAR scale, we measured reading psychophysics and sickness symptoms in a controlled within-subjects study. We find that reading performance and user comfort degrade exponentially as resolution drops below 0 logMAR (normal visual acuity). Notably, our results reveal 0 logMAR as a key physiological tipping point: resolutions better than this threshold yield naked-eye-level performance with minimal sickness, whereas poorer resolutions trigger rapid, non-linear increases in nausea and oculomotor strain. These findings suggest that the cognitive and perceptual effort required to resolve blurry text directly compromises user comfort, establishing human-eye resolution as a critical baseline for the design of future ergonomic XR systems.
\end{abstract}


\begin{bibunit}[unsrtnat]

\section{Introduction}

\maketitle
\label{sec:introduction}
Extended reality (XR) head-mounted displays (HMDs), including virtual reality (VR) and video see-through (VST; passthrough), are increasingly framed as general-purpose computing platforms rather than entertainment devices \cite{vasarainen2021systematic, 10.1145/3691573.3691599}. Productivity tasks such as virtual desktops, in-HMD web browsing, document editing, messaging, coding, and reading physical media via VST rely on two prerequisites: readable text and user comfort. Even small deficits in either depress throughput, raise error rates, shorten sessions, and ultimately drive task abandonment \cite{grout2015reading, 10.1145/3334480.3382920, 9123091, desouza2024vsteval}. Comfort is further constrained by simulator sickness (a cluster of nausea, oculomotor strain, and disorientation) linked to visual–vestibular conflicts, motion-to-photon latency and reprojection, inconsistent optic flow, and accommodation–vergence stress \cite{Stauffert2020Latency,Palmisano2020DVP,Hoffman2008VAC,Shibata2011Zone}. Even mild symptoms degrade attention and reading efficiency and reduce usable session length, as captured by measurements such as the simulator sickness questionnaire (SSQ) \cite{souchet2022measuring, Erickson2020}. Therefore, it is important to seek principled, user-centered measures that relate XR system characteristics to functional reading performance and comfort across VR and VST.

A central reason for these problems is limited \emph{effective} resolution (the clarity that reaches the eye after the full display–optics–rendering pipeline). Conventional specifications such as panel resolution in width $\times$ height, pixels per inch (PPI), or pixels per degree (PPD) are useful engineering descriptors, but they do not fully capture the end-to-end visual clarity that users experience under HMDs \cite{lu2023display}. Optical aberrations, render resolution, and image processing can all alter what the eye ultimately resolves. Traditional vision science metrics and display-based visual assessments are increasingly driving the establishment of visual perception standards for HMDs. Recent XR clarity metrics, such as omnidirectional virtual visual acuity (OVVA), address this by expressing effective resolution directly in the logarithm of the minimum angle of resolution (logMAR), a vision-centric unit where 0 logMAR approximates normal human visual acuity \cite{ovva}. Extending this pursuit of fidelity, Ashraf et al. \cite{ashraf2025resolution} have recently investigated the ultimate resolution limits of the human eye, providing quantitative benchmarks to determine the threshold at which further display resolution improvements yield no perceivable benefit. Framing resolution in logMAR allows us to investigate human performance as a function of perceived clarity, independent of the particular combination of display panel, optics, and rendering pipeline. This perspective is especially valuable for HMDs, where clarity can vary substantially across devices and viewing conditions.

Reading is a foundational task in human–computer interaction (HCI) and a sensitive probe of visual performance \cite{monk2014fundamentals, Clinton2019}. The reading performance can be measured by psychophysical methods such as the reading acuity test \cite{legge2006psychophysics}. For example, the Minnesota low vision reading chart (MNREAD acuity chart) is widely used to characterize reading performance via maximum reading speed (MRS), critical print size (CPS), reading acuity (RA), and the reading accessibility index (ACC) \cite{legge1989MNREAD,Mansfield1993MNREAD,Rubin2013Measuring,calabrese2016ACC}. However, existing implementations are not designed to operate across VR and VST, and do not support multiple languages with comparable stimuli, tightly controlled, repeatable levels of effective resolution \cite{weir2023see, han2017cread, calabrese2018comparing}. This limits our ability to generate generalizable guidance for XR systems intended for everyday reading.

We address this gap with \emph{XR-Read}, a language-switchable reading acuity test that precisely controls effective resolution (in logMAR) and runs natively in VR and on smartphones for VST conditions. Using XR-Read, we conduct a within-subjects study in which participants read English and Chinese sentences at four effective resolutions (0.0, 0.2, 0.4, 0.6 logMAR). We record both the standard reading performance (MRS, CPS, RA, ACC) and symptoms of simulator sickness.

Our results show that across languages and display modes, degrading effective resolution systematically reduces MARS, increases CPS and RA thresholds, lowers the ACC, and elevates simulator sickness symptoms. These convergent effects indicate that effective resolution is a primary driver of both reading performance and comfort in HMDs. Building on these findings, we provide generalized reference curves that map logMAR to reading metrics and simulator sickness symptoms, enabling designers and engineers to set clarity targets, choose rendering budgets, and select user interface (UI) typography that preserves readability in everyday use.

In summary, this paper makes three contributions:
\begin{itemize}
  \item \textbf{A cross-platform, language-switchable reading acuity test} for VR and VST that precisely controls effective resolution in logMAR. The source code is available for download (\url{https://github.com/Chaosikaros/XR-Read})
  \item \textbf{A within-subjects evaluation} quantifying how effective resolution (0.0, 0.2, 0.4, 0.6 logMAR) affects reading performance (MRS, CPS, RA, ACC) and simulator sickness symptoms across display modes and languages.
\item \textbf{Generalized reference curves and actionable guidance} linking logMAR to reading performance and comfort, \textbf{establishing \textit{normal visual acuity (0 logMAR)} as a practical operating threshold} for sustained, low-sickness reading in XR, plus a repeatable protocol for resolution-dependent readability assessment.
\end{itemize}

\section{Related Work} \label{sec:related_work}

\subsection{Visual Clarity in XR HMDs}
Traditional, pixel-centric specifications such as panel resolution, subpixel layout, and PPD do not predict the visual detail that users actually resolve once optics, rendering, and post-processing are in the loop. In practice, lens modulation transfer function (MTF), pupil swim, chromatic aberration correction, temporal upsampling, anti-aliasing, and sharpening/denoising each reshape the modulation spectrum that reaches the eye; the render resolution scale then acts as a hard bandwidth limit on top of those effects \cite{lu2023display}. Recent user-centric metrics, therefore, shift from nominal pixels to \emph{effective resolution} in vision-science units (e.g., logMAR). OVVA proposes this idea by turning a visual acuity chart into an in-HMD psychophysical measurement and reporting clarity directly in logMAR across view directions, enabling fair comparisons across devices and pipelines that differ in optics and render scale \cite{ovva}. A key result from OVVA is that PPD is an unreliable metric for perceived clarity: on a fixed HMD, varying render resolution alone can change measured logMAR, and a lower-PPD device rendered at a higher render resolution can outperform a higher-PPD device rendered below native. This explains why nominal pixel metrics routinely overestimate what users can resolve.

VST adds an imaging stage—capture, image signal processor (ISP), reprojection, and display—that further affects effective resolution and can introduce temporal artifacts relative to optical see-through (OST). Camera sensor, demosaicing, rolling shutter, tone mapping, and ISP sharpening/denoising all interact with HMD optics and the chosen render scale; the net effect is typically a contrast-loss and detail roll-off that show up as worse logMAR, even when nominal panel specs are unchanged \cite{PassthroughPlus, Pohl2013, Carkeet2021}. Because the VST stream is reprojected into eye space, any error in calibration or pose prediction can degrade high-frequency features over time, again depressing effective resolution \cite{NeuralPassthrough}. 
These results argue for clarity metrics and typography guidelines that are defined in logMAR at the eye rather than in panel pixels, and for reporting the render resolution scales that produced those logMAR values.

\subsection{Reading Performance in XR HMDs}
The MNREAD reading visual acuity test is a well-established tool for measuring reading performance via MRS, CPS, and RA \cite{legge1989MNREAD}. Later, Calabr\`{e}se et al.\ introduced the ACC as a single summary measure of reading accessibility across print sizes, normalized to normal vision \cite{calabrese2016ACC}. This index captures both acuity and fluency, and has been proposed for evaluating interventions or devices in low vision and aging populations.

To enable cross-language research, MNREAD-style charts have been created in multiple scripts. Han et al.\ developed the C-READ chart for simplified Chinese, verifying that RA, CPS, and MRS outcomes in young readers aligned with expectations and correlated with standard text reading \cite{han2017cread}. Cheung et al.\ further developed and validated a Chinese reading chart for children, preserving MNREAD psychometrics and confirming reliability across versions \cite{cheung2015chinesereading}. Such tools allow comparisons of reading performance across languages and age groups on a common logMAR scale. However, physical MNREAD and the current digital iPad version do not have enough sentences for massive repeated measures and only support a few languages \cite{mansfield2019extending}.

Within XR, researchers have begun linking reading performance to HMD display parameters. Kilpel\"ainen and H\"akkinen compared legibility across HMDs and demonstrated that perception-first methods reveal clear device differences, arguing that XR typography and rendering requirements should be based on what users can actually resolve rather than engineering specs \cite{kilpelainen2023xrlegibility}. Manipulating angular text size and resolution also shows predictable degradations in reading speed and the smallest readable size as effective resolution of HMDs worsens \cite{novotny2024readingvr}. However, these studies did not compare against VST, which is more susceptible to legibility issues \cite{10972690}. Moreover, they did not consider and control render resolution when comparing effective resolutions. In this case, the results from default render resolution can be misleading since default render resolution may fail to leverage the nominal PPD according to OVVA \cite{ovva}.

Existing XR reading studies typically: evaluate a single language or a single display modality, limiting generalization; use hardware comparisons that confound optics, rendering, and latency with clarity; or lack an explicit link from logMAR clarity to reading performance, making it hard to set actionable typography or rendering budgets. Therefore, we introduce XR-Read, a language-switchable reading acuity test that runs natively in VR and supports VST via a smartphone display, allowing participants to read matched multiple language sentences under the different clarity levels. 

Moreover, we isolate \emph{effective resolution} as the primary factor by directly controlling clarity in logMAR while holding other variables constant on a modern, high-performance HMD (Varjo XR-4 Focal Edition with 55 PPD VR, 51 PPD VST). Using the same device and task across modalities, we induce stepwise degradations (0.0, 0.2, 0.4, 0.6 logMAR) and quantify their impact on functional reading performance and comfort. Unlike most prior work, we explicitly manipulate and report render resolution scales so that logMAR reflects true end-to-end clarity rather than default render resolution. By manipulating only effective resolution and collecting MRS, CPS, RA, and ACC within-subjects across modalities and languages, we derive generalized reference curves mapping logMAR to functional reading performance. 

\subsection{Simulator Sickness in XR HMDs}
Comfort is as essential as legibility for sustained XR use. Simulator sickness (or cybersickness) refers to nausea, oculomotor strain, disorientation, and dizziness that occur during immersion. It is commonly assessed with the Simulator Sickness Questionnaire (SSQ) \cite{kennedy1993ssq}. Even mild simulator sickness degrades productivity by forcing breaks and shortening session lengths \cite{souchet2022measuring}.

Extensive research shows that simulator sickness depends on both system and content factors. Saredakis et al.’s meta-analysis confirmed systematic effects of FoV, motion, and exposure duration on sickness scores \cite{saredakis2020meta}. In VST, camera quality, latency, and visual fidelity also affect comfort. De Souza and Tartz found that participants reported device-dependent differences in nausea and visual disturbances \cite{desouza2024vsteval}. 

While many studies manipulate motion, FoV, or frame rate, fewer quantify how reading performance vary with different VR or VST environments, none of the related studies examine whether effective-resolution-driven legibility losses vary with simulator sickness symptoms across both VR and VST under tightly controlled environments \cite{kilpelainen2023xrlegibility, Fernandes2016, wang2023effect}.
We measure simulator sickness alongside reading performance while varying only the effective resolution on a single HMD. This yields joint logMAR, reading metrics, and simulator sickness curves that expose shared operating regions where readability and comfort are both preserved. By decoupling clarity from other system factors, our results provide concrete, effective resolution targets (in logMAR) for XR text rendering and UI design, and practical guidance for choosing type scales and rendering budgets that maintain both throughput and comfort in VR and VST.

\subsection{Research Questions}
According to our literature review and the gap we identified, our study is guided by three research questions (RQs):
\begin{itemize}
\item \textbf{RQ1:} How does effective resolution (logMAR) affect functional reading performance (MRS, CPS, RA, ACC) in HMDs?
\item \textbf{RQ2:} Do these effects generalize across display modes (VR vs. VST) and languages (English vs. Chinese)?
\item \textbf{RQ3:} How does effective resolution relate to self-reported simulator sickness, and can we derive practical operating points for comfort?
\end{itemize}

\section{XR-Read: Reading Acuity Test for VR and VST}
In this section, we introduce the metric definitions of the reading acuity test, its original design, and our improved digital one, followed by a user study conducted with the digital one and its results and discussion.

\subsection{Metric Definitions} 
In vision science, print size is characterized not by its physical dimensions 
alone but by the \emph{visual angle} it subtends at the eye. 
The visual angle represents how large an object appears on the retina 
and provides a scale-invariant measure that accounts for both the physical 
size of the text and the viewing distance. 
This makes visual angle a natural unit for describing legibility and acuity.

If $h$ denotes the x-height (the height of a lower-case ``x'') 
and $d$ the viewing distance, the visual angle $\theta_{x}$ 
subtended by the print is given by

\begin{equation} \label{eq:theta}
\theta_{x} \;=\; 2 \arctan\!\left( \frac{h}{2d} \right).
\end{equation}

The \emph{logMAR print size} (logarithm of the Minimum Angle of Resolution) 
is defined as

\begin{equation} \label{eq:logmar}
S \;=\; \log_{10}\!\left( \frac{\theta_{x}}{5 \,\text{arc min}} \right),
\end{equation}

where $S$ denotes the print size in logMAR units, $\theta_{x}$ is the visual angle 
(in arc minutes (arc min)) subtended by the x-height, and $5$ arc min is the reference angle 
corresponding to standard visual acuity. 
Thus, the logMAR value quantifies how the size of printed text compares 
to the resolution limit of normal vision.

The relationship between logMAR visual acuity and decimal visual acuity is given by

\begin{equation} \label{eq:decimal}
S \;=\; -\log_{10}\!\left( \text{Decimal acuity} \right),
\end{equation}

By convention, \emph{normal visual acuity} is defined as $S = 0$, 
which corresponds to a decimal acuity of $1.0$, also expressed 
as 20/20 vision in Snellen notation \cite{holladay2004visual}.

Let $\mathrm{RS}(S)$ denote reading speed in words per minute (wpm) at print size $S$. For a sentence read in $t$ seconds with $e$ word errors (MNREAD uses 10 (in average) words per sentence, C-READ uses 12-character simplified Chinese per sentence \cite{calabrese2018comparing, han2017cread}), the sentence-level reading speed is
\begin{equation} \label{eq:RS}
\mathrm{RS} \;=\; \frac{60\,(10 - e)}{t}\ \text{wpm}.
\end{equation}

MRS is the asymptotic (plateau) reading speed achieved at sufficiently large print sizes; operationally, it is estimated from the plateau of the fitted MNREAD curve:
\begin{equation} \label{eq:MRS}
\mathrm{MRS} \;=\; \lim_{S\to \text{large}}\ \mathrm{RS}(S).
\end{equation}
Higher values indicate better performance. In young, normally sighted adults, typical values are around $200$\, wpm for both native English and Chinese speakers \cite{calabrese2018comparing, cheng2023using}, remaining relatively stable through early adulthood before gradually declining with age.

CPS is defined as the smallest print size \(S\) (in logMAR) at which reading speed remains near the maximum. Operationally, CPS is often taken as the smallest \(S\) such that the reading speed \(\mathrm{RS}(S)\) is at least a fraction \(p\) of the maximum reading speed (MRS). Many studies use \(p=0.90\), while values in the range \(p\in[0.80,1.00]\) are reported to balance reliability versus strictness \cite{baskaran2019scoring,xiong2018reading2}:
\begin{equation}\label{eq:CPS}
\mathrm{CPS}(p) \;=\; \min\bigl\{\, S \;:\; \mathrm{RS}(S) \,\ge\, p\,\mathrm{MRS} \,\bigr\}, 
\qquad p\in[0.80,1.00].
\end{equation}
Lower CPS indicates better performance because it means smaller print can be read at (nearly) maximum speed. In normally sighted young adults, CPS is typically near \(0.1\)\,logMAR, and increases with age or visual impairment \cite{calabrese2016baseline,calabrese2018comparing}. 
As a supplement, a standard deviation (SD) method definition sets CPS as the smallest size on the reading-speed plateau whose mean is significantly (e.g., \(\ge 1.96\) SD) faster than the speeds at smaller print sizes; MRS is the plateau mean \cite{baskaran2019scoring}.

RA is the print-size threshold (in logMAR) that a person can read without making significant errors:
\begin{equation} \label{eq:RA}
\mathrm{RA} \;=\; S_{\text{smallest readable}} \;+\; 0.01 \times e_{\text{(at that size)}}.
\end{equation}
Lower RA is better (finer acuity). In young, normally sighted observers, RA commonly falls around $-0.2$ to $-0.15$ logMAR, with higher (worse) values in older adults or those with low vision \cite{calabrese2016baseline, calabrese2018comparing}.

ACC summarizes access to everyday text sizes. It is computed as the mean reading speed across the ten largest print sizes, normalized by $200$\, wpm (a young-normal reference):
\begin{equation} \label{eq:ACC}
\mathrm{ACC} \;=\; \frac{1}{10}\sum_{i=1}^{10}\frac{\mathrm{RS}(S_i)}{200}.
\end{equation}
Higher ACC is better. By construction, $\mathrm{ACC}\approx 1.0$ for young, commonly sighted readers; values greater than $1$ reflect above-reference access, whereas values approaching $0$ indicate severely limited access to common print \cite{calabrese2016ACC}.

The MNREAD data consist of pairs $\bigl(S_i,\mathrm{RS}_i\bigr)$ across descending print sizes. Because $S$ is an angular measure, any change in viewing distance $d$ results in a uniform offset of all print sizes. If two sessions are administered at distances $d$ and $d'$, their size coordinates obey
\begin{equation}\label{eq:distance-shift}
S(d') \;=\; S(d) \;+\; \log_{10}\!\left(\frac{d'}{d}\right),
\end{equation}
so that optional standardization to a 40\,cm reference is obtained by $S_{40} = S(d) + \log_{10}(0.40/d)$. No adjustment to $\mathrm{RS}$ is necessary because reading speed is distance-invariant within the MNREAD paradigm when $S$ is defined angularly.

The original MNREAD testing begins at a large size and proceeds in 0.1 logMAR steps toward smaller sizes ($-0.5$ to $1.3$; 19 sentences) until the participant can no longer produce any words for a given sentence. The examiner instructs the participant to read aloud as quickly and accurately as possible, and to guess when uncertain. Viewing distance is measured and maintained; a chinrest or rigid measuring tape improves stability, and the final distance is recorded so that $S$ reflects true angular print size. For each sentence, the examiner captures the elapsed time $t$, tallies $e$, and computes $\mathrm{RS}$ via \eqref{RS}.

\begin{figure}[!ht]
 \centering\includegraphics[width=\columnwidth]{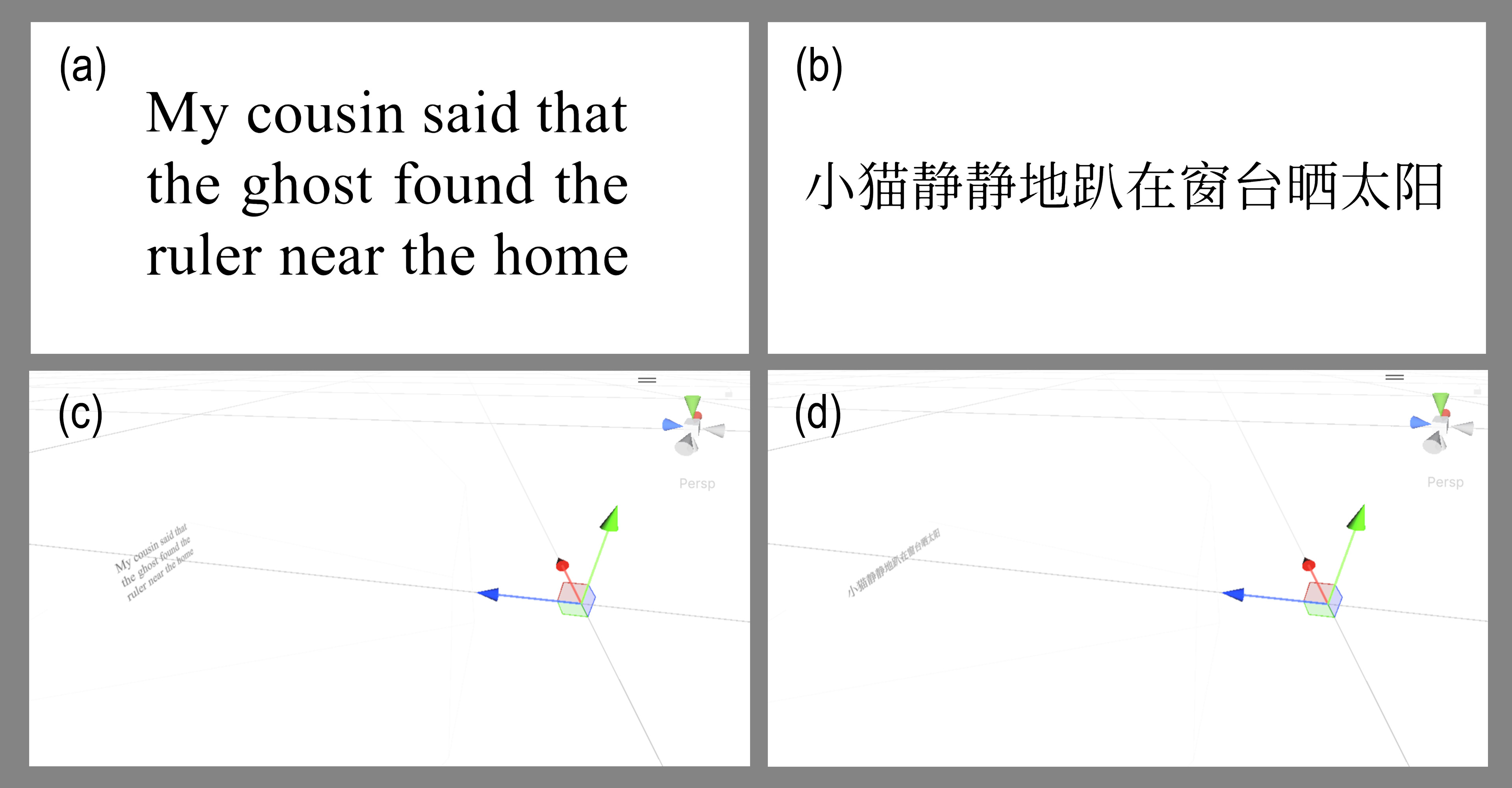}
 \caption{Screenshots of English ((a) and (c)) and Chinese ((b) and (d)) sentences from XR-Read. The first row shows the 2D XR-Read captured on a smartphone. The second row shows 3D XR-Read views captured in the Unity editor, where the blue arrow indicates the direction of the eye camera. }
 \label{fig:screenshots}
\end{figure}

\begin{figure}[!ht]
 \centering\includegraphics[width=\columnwidth]{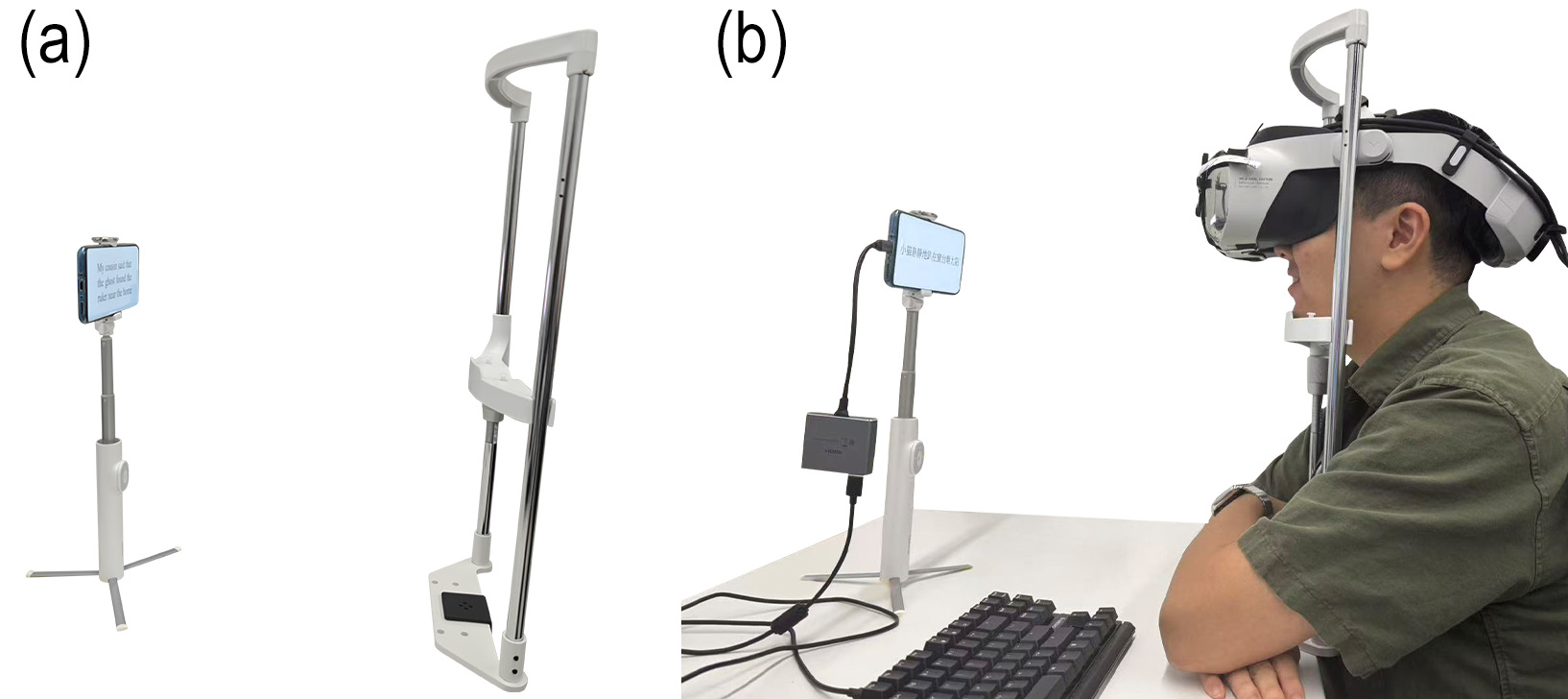}
 \caption{Pictures of the 2D XR-Read on a smartphone and the chin rest. (a) and (b) show the experiment setup without and with a participant wearing the Varjo XR-4 Focal Edition.}
 \label{fig:photos}
\end{figure}

\begin{figure}[!ht]
 \centering\includegraphics[width=\columnwidth]{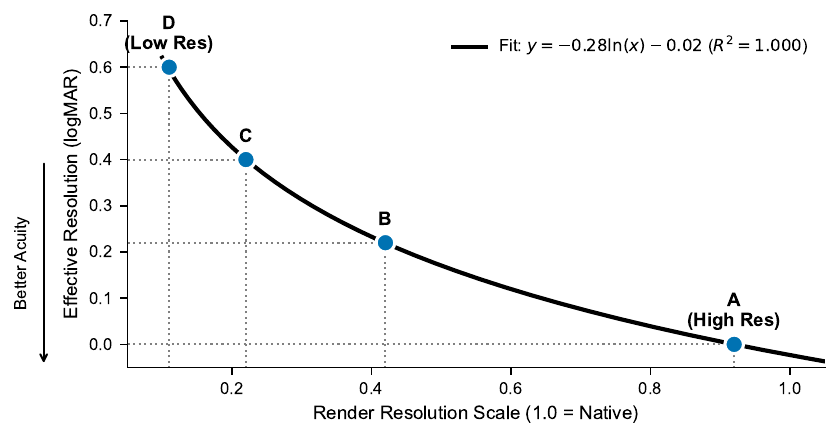}
 \caption{Mapping between render resolution scale and effective resolution in VR. The logarithmic regression $y=-0.28\ln(x)-0.02$ describes the relationship between render resolution scale $x$ and visual acuity (logMAR) $y$, measured with the OVVA test. Red markers (A--D) indicate the four target acuity levels (A = 0.00, B = 0.22, C = 0.40, D = 0.60 logMAR) and their corresponding scales ($x \approx $ 0.92, 0.42, 0.22, 0.11), with dashed lines projected to the axes.}
 \label{fig:resolution_curve}
\end{figure}

\begin{figure*}[!ht]
 \centering\includegraphics[width=\textwidth]{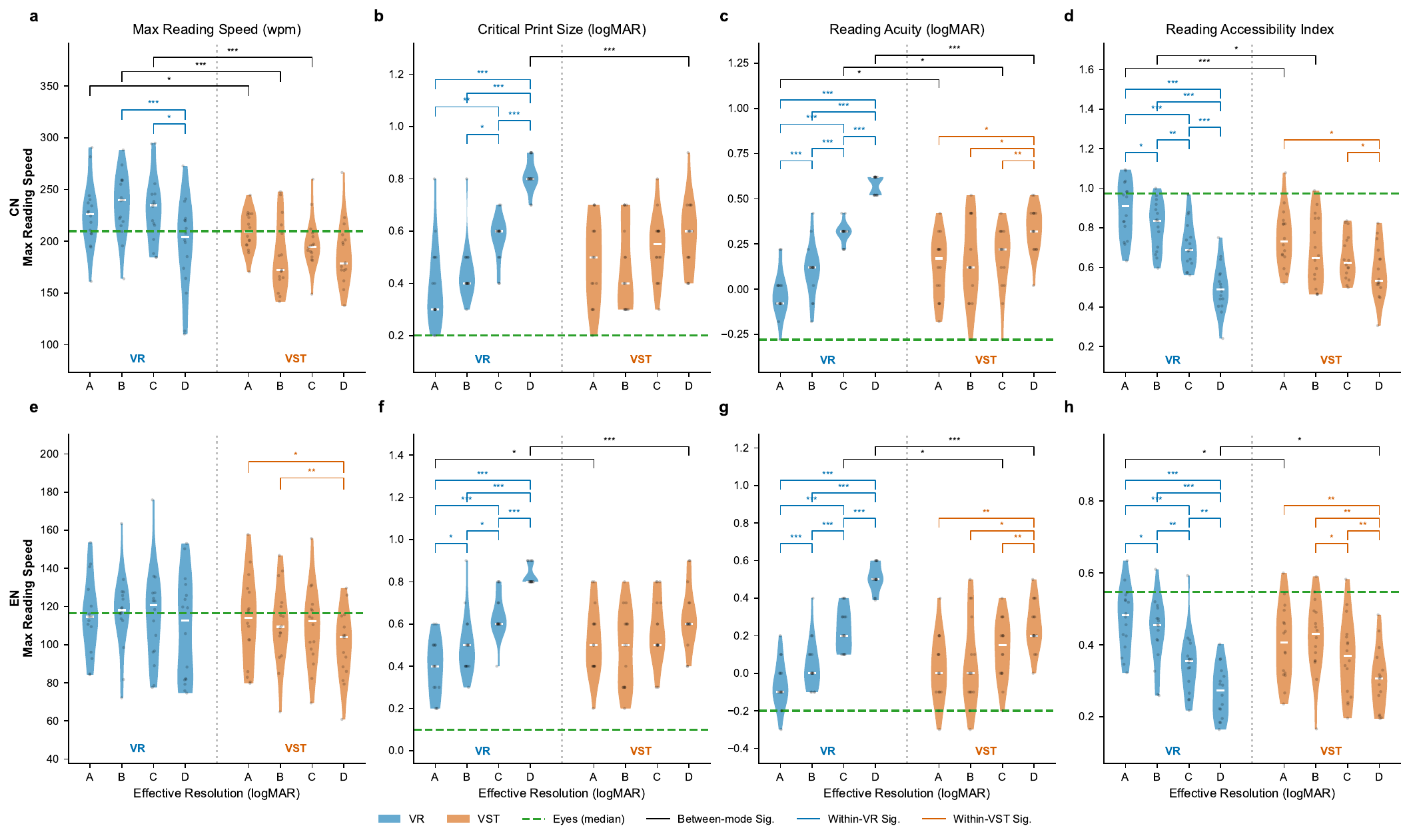}
 \caption{Violin plots with post-hoc results of XR-Read. `*' to `***' represent Bonferroni-adjusted significant differences at `.05', `.01', `.001' level.}
 \label{fig:MNREAD}
\end{figure*}

\begin{table}[!ht]
\centering
\begin{tabular}{lccccc}
\hline
\toprule
\textbf{Display} & \textbf{Metric} & \textbf{$\chi^2$} & \textbf{df} & \textbf{p-value} & \textbf{$W$} \\
\midrule
\multirow{8}{*}{VST} & ACC (CN) & 15.4 & 3 & \highlightcell{0.002} & 0.320 \\
  & ACC (EN) & 18.7 & 3 & \highlightcell{$<$0.001} & 0.389 \\
  & CPS (CN) & 7.9 & 3 & \highlightcell{0.048} & 0.164 \\
  & CPS (EN) & 9.2 & 3 & \highlightcell{0.026} & 0.192 \\
  & MRS (CN) & 8.0 & 3 & \highlightcell{0.045} & 0.167 \\
  & MRS (EN) & 9.5 & 3 & \highlightcell{0.023} & 0.198 \\
  & RA (CN) & 15.6 & 3 & \highlightcell{0.001} & 0.324 \\
  & RA (EN) & 21.4 & 3 & \highlightcell{$<$0.001} & 0.446 \\
\midrule
\multirow{8}{*}{VR} & ACC (CN) & 37.3 & 3 & \highlightcell{$<$0.001} & 0.778 \\
  & ACC (EN) & 39.2 & 3 & \highlightcell{$<$0.001} & 0.817 \\
  & CPS (CN) & 36.1 & 3 & \highlightcell{$<$0.001} & 0.752 \\
  & CPS (EN) & 40.2 & 3 & \highlightcell{$<$0.001} & 0.838 \\
  & MRS (CN) & 18.5 & 3 & \highlightcell{$<$0.001} & 0.386 \\
  & MRS (EN) & 2.5 & 3 & 0.480 & 0.052 \\
  & RA (CN) & 44.4 & 3 & \highlightcell{$<$0.001} & 0.926 \\
  & RA (EN) & 45.5 & 3 & \highlightcell{$<$0.001} & 0.947 \\
\bottomrule
\end{tabular}
\caption{Friedman tests for XR-Read across the four \textbf{Effective Resolution} levels (A, B, C, D) within each \textbf{Display} (VST, VR) for each language (CN, EN). Significant \highlighttextbf{p-values} are highlighted.}
\label{tab:MNREAD_Friedman}
\end{table}

\begin{table}[!ht]
\centering
\begin{tabular}{p{1cm} p{1cm} c c c c}
\hline
\toprule
& & \multicolumn{2}{c}{\textbf{VST}} & \multicolumn{2}{c}{\textbf{VR}} \\
\cmidrule(lr){3-4} \cmidrule(lr){5-6}
\textbf{Metric} & \textbf{Pair} & \textbf{$W$} & \textbf{p-value} & \textbf{$W$} & \textbf{p-value} \\
\midrule
\multirow{6}{*}{MRS} & A vs.B & 24.0 & 0.107 & 32.0 & 0.196 \\
  & A vs.C & 28.0 & 0.155 & 44.0 & 0.462 \\
  & A vs.D & 22.0 & 0.093 & 23.0 & 0.073 \\
  & B vs.C & 34.0 & 0.250 & 55.0 & 0.528 \\
  & B vs.D & 54.0 & 0.751 & 1.0 & \highlightcell{$<$0.001***} \\
  & C vs.D & 50.0 & 0.751 & 12.0 & \highlightcell{0.011*} \\
\midrule
\multirow{6}{*}{CPS} & A vs.B & 27.5 & 0.902 & 31.0 & 0.107 \\
  & A vs.C & 24.5 & 0.902 & 8.0 & \highlightcell{0.009**} \\
  & A vs.D & 20.5 & 0.198 & 0.0 & \highlightcell{$<$0.001***} \\
  & B vs.C & 9.0 & 0.081 & 15.0 & \highlightcell{0.033*} \\
  & B vs.D & 12.5 & 0.064 & 0.0 & \highlightcell{$<$0.001***} \\
  & C vs.D & 13.0 & 0.902 & 0.0 & \highlightcell{$<$0.001***} \\
\midrule
\multirow{6}{*}{RA} & A vs.B & 30.0 & 0.916 & 5.0 & \highlightcell{$<$0.001***} \\
  & A vs.C & 26.0 & 1.000 & 0.0 & \highlightcell{$<$0.001***} \\
  & A vs.D & 10.0 & \highlightcell{0.013*} & 0.0 & \highlightcell{$<$0.001***} \\
  & B vs.C & 59.0 & 1.000 & 2.5 & \highlightcell{$<$0.001***} \\
  & B vs.D & 7.0 & \highlightcell{0.019*} & 0.0 & \highlightcell{$<$0.001***} \\
  & C vs.D & 0.0 & \highlightcell{0.003**} & 0.0 & \highlightcell{$<$0.001***} \\
\midrule
\multirow{6}{*}{ACC} & A vs.B & 39.0 & 0.288 & 20.0 & \highlightcell{0.011*} \\
  & A vs.C & 25.0 & 0.086 & 5.0 & \highlightcell{$<$0.001***} \\
  & A vs.D & 11.0 & \highlightcell{0.010*} & 0.0 & \highlightcell{$<$0.001***} \\
  & B vs.C & 52.0 & 0.433 & 7.0 & \highlightcell{0.001**} \\
  & B vs.D & 24.0 & 0.086 & 0.0 & \highlightcell{$<$0.001***} \\
  & C vs.D & 16.0 & \highlightcell{0.026*} & 1.0 & \highlightcell{$<$0.001***} \\
\bottomrule
\end{tabular}
\caption{Wilcoxon signed-rank tests with Bonferroni correction for XR-Read within \textbf{Display} across all \textbf{Effective Resolution} pairs (CN). Green p-values with `*' to `***' indicate adjusted significance at .05, .01, and .001 levels, respectively.}
\label{tab:MNREAD_Posthoc_resolution_CN}
\end{table}

\begin{table}[!ht]
\centering
\begin{tabular}{p{1cm} p{1cm} c c c c}
\hline
\toprule
& & \multicolumn{2}{c}{\textbf{VST}} & \multicolumn{2}{c}{\textbf{VR}} \\
\cmidrule(lr){3-4} \cmidrule(lr){5-6}
\textbf{Metric} & \textbf{Pair} & \textbf{$W$} & \textbf{p-value} & \textbf{$W$} & \textbf{p-value} \\
\midrule
\multirow{6}{*}{MRS} & A vs.B & 50.0 & 0.549 & 60.0 & 1.000 \\
  & A vs.C & 39.0 & 0.432 & 67.0 & 1.000 \\
  & A vs.D & 18.0 & \highlightcell{0.038*} & 43.0 & 1.000 \\
  & B vs.C & 46.0 & 0.549 & 63.0 & 1.000 \\
  & B vs.D & 8.0 & \highlightcell{0.005**} & 33.0 & 0.443 \\
  & C vs.D & 21.0 & 0.052 & 45.0 & 1.000 \\
\midrule
\multirow{6}{*}{CPS} & A vs.B & 32.0 & 0.751 & 6.5 & \highlightcell{0.027*} \\
  & A vs.C & 50.5 & 0.952 & 0.0 & \highlightcell{$<$0.001***} \\
  & A vs.D & 7.0 & 0.239 & 0.0 & \highlightcell{$<$0.001***} \\
  & B vs.C & 22.0 & 0.239 & 11.0 & \highlightcell{0.027*} \\
  & B vs.D & 12.0 & 0.051 & 0.0 & \highlightcell{$<$0.001***} \\
  & C vs.D & 16.5 & 0.239 & 0.0 & \highlightcell{$<$0.001***} \\
\midrule
\multirow{6}{*}{RA} & A vs.B & 46.5 & 0.761 & 0.0 & \highlightcell{$<$0.001***} \\
  & A vs.C & 19.5 & 0.126 & 0.0 & \highlightcell{$<$0.001***} \\
  & A vs.D & 0.0 & \highlightcell{0.003**} & 0.0 & \highlightcell{$<$0.001***} \\
  & B vs.C & 17.0 & 0.185 & 2.0 & \highlightcell{$<$0.001***} \\
  & B vs.D & 9.0 & \highlightcell{0.032*} & 0.0 & \highlightcell{$<$0.001***} \\
  & C vs.D & 1.5 & \highlightcell{0.007**} & 0.0 & \highlightcell{$<$0.001***} \\
\midrule
\multirow{6}{*}{ACC} & A vs.B & 53.0 & 0.464 & 22.0 & \highlightcell{0.016*} \\
  & A vs.C & 41.0 & 0.351 & 2.0 & \highlightcell{$<$0.001***} \\
  & A vs.D & 7.0 & \highlightcell{0.003**} & 0.0 & \highlightcell{$<$0.001***} \\
  & B vs.C & 14.0 & \highlightcell{0.010*} & 6.0 & \highlightcell{0.001**} \\
  & B vs.D & 5.0 & \highlightcell{0.002**} & 0.0 & \highlightcell{$<$0.001***} \\
  & C vs.D & 11.0 & \highlightcell{0.007**} & 10.0 & \highlightcell{0.003**} \\
\bottomrule
\end{tabular}
\caption{Wilcoxon signed-rank tests with Bonferroni correction for XR-Read within \textbf{Display} across all \textbf{Effective Resolution} pairs (EN). Green p-values with `*' to `***' indicate adjusted significance at .05, .01, and .001 levels, respectively.}
\label{tab:MNREAD_Posthoc_resolution_EN}
\end{table}

\begin{table*}[!ht]
\centering
\setlength{\tabcolsep}{4pt}
\renewcommand{\arraystretch}{1.1}
\begin{tabular}{
  p{1.1cm} p{1.6cm} *{4}{c}
  @{\hspace{0.9cm}}
  p{1.6cm} *{4}{c}
}
\toprule
\multicolumn{6}{c}{\textbf{CN vs. EN}} &
\multicolumn{5}{c}{\textbf{VST vs. VR}}\\
\cmidrule(lr){1-6}\cmidrule(lr){7-11}
\multicolumn{1}{c}{\textbf{Metric}} &
\multicolumn{1}{c}{\textbf{Resolution}} &
\multicolumn{2}{c}{\textbf{VST}} &
\multicolumn{2}{c}{\textbf{VR}} &
\multicolumn{1}{c}{\textbf{Resolution}} &
\multicolumn{2}{c}{\textbf{CN}} &
\multicolumn{2}{c}{\textbf{EN}} \\
\cmidrule(lr){3-4}\cmidrule(lr){5-6}\cmidrule(lr){8-9}\cmidrule(lr){10-11}
& & \textbf{$W$} & \textbf{p-value} & \textbf{$W$} & \textbf{p-value}
& & \textbf{$W$} & \textbf{p-value} & \textbf{$W$} & \textbf{p-value} \\
\midrule
\multirow{4}{*}{MRS} & A & 0.0 & \highlightcell{$<$0.001***} & 0.0 & \highlightcell{$<$0.001***} & A & 20.0 & \highlightcell{0.022*} & 45.0 & 0.296 \\
  & B & 0.0 & \highlightcell{$<$0.001***} & 0.0 & \highlightcell{$<$0.001***} & B & 3.0 & \highlightcell{$<$0.001***} & 35.0 & 0.296 \\
  & C & 0.0 & \highlightcell{$<$0.001***} & 0.0 & \highlightcell{$<$0.001***} & C & 3.0 & \highlightcell{$<$0.001***} & 33.0 & 0.296 \\
  & D & 0.0 & \highlightcell{$<$0.001***} & 3.0 & \highlightcell{$<$0.001***} & D & 59.0 & 0.669 & 37.0 & 0.296 \\
\midrule
\multirow{4}{*}{CPS} & A & 16.5 & 0.554 & 23.0 & 1 & A & 17.5 & 0.330 & 9.5 & \highlightcell{0.031*} \\
  & B & 26.0 & 1 & 9.0 & 0.451 & B & 43.5 & 0.946 & 46.0 & 0.715 \\
  & C & 22.0 & 1 & 2.0 & 0.188 & C & 18.0 & 0.750 & 18.0 & 0.220 \\
  & D & 24.0 & 0.732 & 5.0 & 1 & D & 0.0 & \highlightcell{$<$0.001***} & 4.0 & \highlightcell{$<$0.001***} \\
\midrule
\multirow{4}{*}{RA} & A & 0.0 & \highlightcell{$<$0.001***} & 26.0 & 0.087 & A & 6.5 & \highlightcell{0.010*} & 23.5 & 0.157 \\
  & B & 20.0 & \highlightcell{0.044*} & 26.0 & 0.087 & B & 26.0 & 0.339 & 32.0 & 0.966 \\
  & C & 30.0 & 0.087 & 11.0 & \highlightcell{0.008**} & C & 5.5 & \highlightcell{0.010*} & 10.0 & \highlightcell{0.031*} \\
  & D & 9.0 & \highlightcell{0.007**} & 10.0 & \highlightcell{0.008**} & D & 0.0 & \highlightcell{$<$0.001***} & 0.0 & \highlightcell{$<$0.001***} \\
\midrule
\multirow{4}{*}{ACC} & A & 0.0 & \highlightcell{$<$0.001***} & 0.0 & \highlightcell{$<$0.001***} & A & 3.0 & \highlightcell{$<$0.001***} & 18.0 & \highlightcell{0.031*} \\
  & B & 0.0 & \highlightcell{$<$0.001***} & 0.0 & \highlightcell{$<$0.001***} & B & 19.0 & \highlightcell{0.028*} & 49.0 & 0.697 \\
  & C & 0.0 & \highlightcell{$<$0.001***} & 0.0 & \highlightcell{$<$0.001***} & C & 34.0 & 0.083 & 56.0 & 0.697 \\
  & D & 0.0 & \highlightcell{$<$0.001***} & 0.0 & \highlightcell{$<$0.001***} & D & 27.0 & 0.067 & 18.0 & \highlightcell{0.031*} \\
\bottomrule
\end{tabular}
\caption{Wilcoxon tests with Bonferroni correction on \textbf{Effective Resolution} and \textbf{Language} of XR-Read. Left: CN vs. EN. Right: VST vs. VR. Highlighted p-values with `*' to `***' indicate adjusted significance at .05, .01, and .001, respectively.}
\label{tab:XRRead_combined_horizontal_simplified_updated}
\end{table*}

\begin{figure*}[!ht]
 \centering\includegraphics[width=\textwidth]{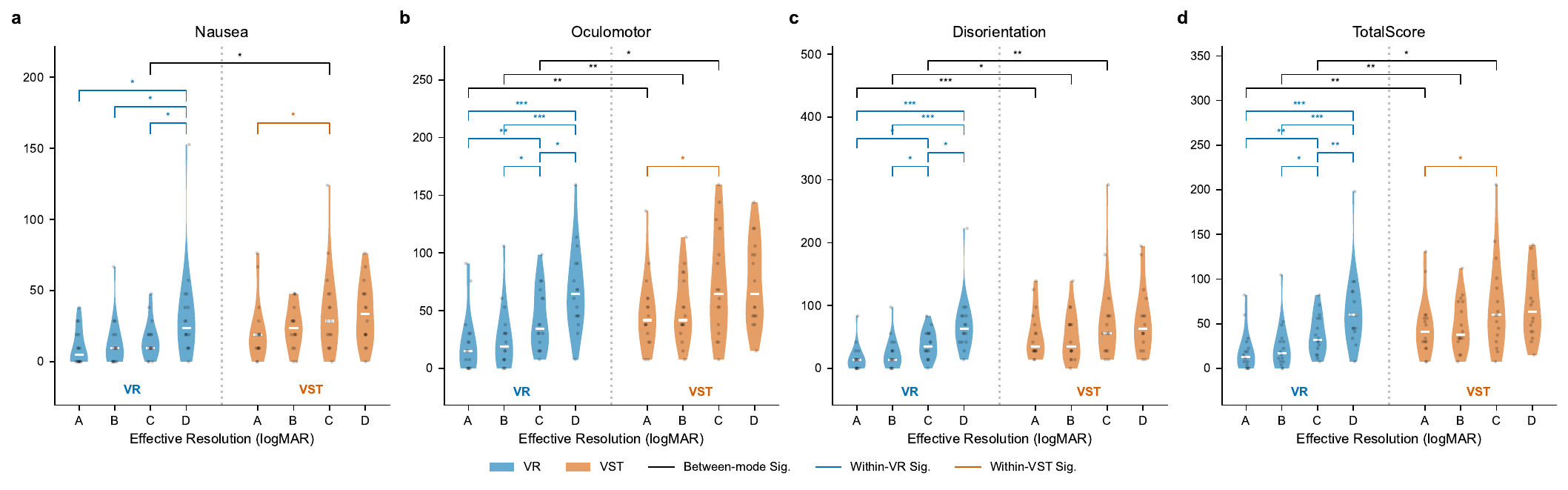}
 \caption{Violin plots with post-hoc results of Simulator Sickness Questionnaire. `*' to `***' represent Bonferroni-adjusted significant differences at `.05', `.01', `.001' level.}
 \label{fig:SSQ}
\end{figure*}

\begin{table}[!ht]
\centering
\begin{tabular}{lccccc}
\hline
\toprule
\textbf{Display} & \textbf{Metric} & \textbf{$\chi^2$} & \textbf{df} & \textbf{p-value} & \textbf{$W$} \\
\midrule
\multirow{4}{*}{VST} & Nausea & 11.2 & 3 & \highlightcell{0.011} & 0.233 \\
 & Oculomotor & 10.7 & 3 & \highlightcell{0.013} & 0.223 \\
 & Disorientation & 5.3 & 3 & 0.15 & 0.111 \\
 & TotalScore & 11.2 & 3 & \highlightcell{0.011} & 0.232 \\
\midrule
\multirow{4}{*}{VR} & Nausea & 20.6 & 3 & \highlightcell{$<$0.001} & 0.429 \\
 & Oculomotor & 28.8 & 3 & \highlightcell{$<$0.001} & 0.600 \\
 & Disorientation & 33.0 & 3 & \highlightcell{$<$0.001} & 0.688 \\
 & TotalScore & 32.2 & 3 & \highlightcell{$<$0.001} & 0.671 \\
\bottomrule
\end{tabular}
\caption{Friedman tests on Simulator Sickness Questionnaire across the four \textbf{Effective Resolution} levels (A, B, C, D) within each \textbf{Display} (VST, VR). Significant \highlighttextbf{p-values} are highlighted.}
\label{tab:SSQ_Friedman}
\end{table}

\begin{table}[!ht]
\centering
\begin{tabular}{p{1.5cm} p{1cm} c c c c}
\hline
\toprule
& & \multicolumn{2}{c}{\textbf{VST}} & \multicolumn{2}{c}{\textbf{VR}} \\
\cmidrule(lr){3-4} \cmidrule(lr){5-6}
\textbf{Metric} & \textbf{Pair} & \textbf{$W$} & \textbf{p-value} & \textbf{$W$} & \textbf{p-value} \\
\midrule
\multirow{6}{*}{Nausea} & A vs.B & 21.0 & 1 & 6.0 & 0.875 \\
 & A vs.C & 0.0 & \highlightcell{0.047*} & 6.0 & 0.328 \\
 & A vs.D & 9.0 & 0.161 & 4.0 & \highlightcell{0.014*} \\
 & B vs.C & 10.5 & 0.422 & 13.0 & 0.875 \\
 & B vs.D & 19.5 & 0.454 & 4.0 & \highlightcell{0.01*} \\
 & C vs.D & 34.5 & 1 & 0.0 & \highlightcell{0.01*} \\
\midrule
\multirow{6}{*}{Oculomotor} & A vs.B & 44.5 & 1 & 17.5 & 0.206 \\
 & A vs.C & 5.5 & \highlightcell{0.041*} & 11.0 & \highlightcell{0.007**} \\
 & A vs.D & 19.0 & 0.09 & 1.0 & \highlightcell{$<$0.001***} \\
 & B vs.C & 21.0 & 0.192 & 10.0 & \highlightcell{0.031*} \\
 & B vs.D & 24.5 & 0.192 & 1.0 & \highlightcell{$<$0.001***} \\
 & C vs.D & 62.0 & 1 & 13.0 & \highlightcell{0.031*} \\
\midrule
\multirow{6}{*}{Disorientation} & A vs.B & 27.0 & 1 & 5.5 & 0.109 \\
 & A vs.C & 16.5 & 0.369 & 12.0 & \highlightcell{0.016*} \\
 & A vs.D & 34.5 & 0.506 & 0.0 & \highlightcell{$<$0.001***} \\
 & B vs.C & 15.0 & 0.196 & 11.0 & \highlightcell{0.027*} \\
 & B vs.D & 15.5 & 0.199 & 1.0 & \highlightcell{$<$0.001***} \\
 & C vs.D & 30.5 & 1 & 9.0 & \highlightcell{0.016*} \\
\midrule
\multirow{6}{*}{TotalScore} & A vs.B & 52.5 & 0.751 & 18.0 & 0.11 \\
 & A vs.C & 8.5 & \highlightcell{0.048*} & 11.0 & \highlightcell{0.007**} \\
 & A vs.D & 17.5 & 0.148 & 0.0 & \highlightcell{$<$0.001***} \\
 & B vs.C & 29.0 & 0.167 & 13.0 & \highlightcell{0.011*} \\
 & B vs.D & 19.5 & 0.167 & 1.0 & \highlightcell{$<$0.001***} \\
 & C vs.D & 49.5 & 0.751 & 9.0 & \highlightcell{0.007**} \\
\bottomrule
\end{tabular}
\caption{Wilcoxon signed-rank tests with Bonferroni correction on Simulator Sickness Questionnaire: all \textbf{Effective Resolution} pairs within each \textbf{Display}. Green p-values with `*' to `***' indicate adjusted significance at .05, .01, and .001 levels, respectively.}
\label{tab:SSQ_Posthoc_resolution}
\end{table}

\begin{table*}[!ht]
\centering
\begin{tabular}{p{1.8cm} p{2.2cm} c c c c c c c c}
\hline
\toprule
& & \multicolumn{2}{c}{\textbf{A}} & \multicolumn{2}{c}{\textbf{B}} & \multicolumn{2}{c}{\textbf{C}} & \multicolumn{2}{c}{\textbf{D}} \\
\cmidrule(lr){3-4} \cmidrule(lr){5-6} \cmidrule(lr){7-8} \cmidrule(lr){9-10}
\textbf{Metric} & \textbf{Pair} & \textbf{$W$} & \textbf{p-value} & \textbf{$W$} & \textbf{p-value} & \textbf{$W$} & \textbf{p-value} & \textbf{$W$} & \textbf{p-value} \\
\midrule
\multirow{1}{*}{Nausea} & VST vs.VR & 5 & 0.059 & 8.5 & 0.129 & 0 & \highlightcell{0.016*} & 31 & 0.569 \\
\multirow{1}{*}{Oculomotor} & VST vs.VR & 10.5 & \highlightcell{0.007**} & 12 & \highlightcell{0.007**} & 8.5 & \highlightcell{0.032*} & 31 & 0.340 \\
\multirow{1}{*}{Disorientation} & VST vs.VR & 1 & \highlightcell{$<$0.001***} & 6 & \highlightcell{0.027*} & 7.5 & \highlightcell{0.009**} & 50 & 0.599 \\
\multirow{1}{*}{TotalScore} & VST vs.VR & 9 & \highlightcell{0.004**} & 11.5 & \highlightcell{0.006**} & 8 & \highlightcell{0.012*} & 49 & 0.348 \\

\bottomrule
\end{tabular}
\caption{Wilcoxon signed-rank tests with Bonferroni correction on Simulator Sickness Questionnaire: all \textbf{Display} pairs within each \textbf{Effective Resolution}. Green p-values with `*' to `***' indicate adjusted significance at .05, .01, and .001 levels, respectively.}
\label{tab:SSQ_Posthoc_display}
\end{table*}

\begin{figure*}[!ht]
 \centering\includegraphics[width=\textwidth]{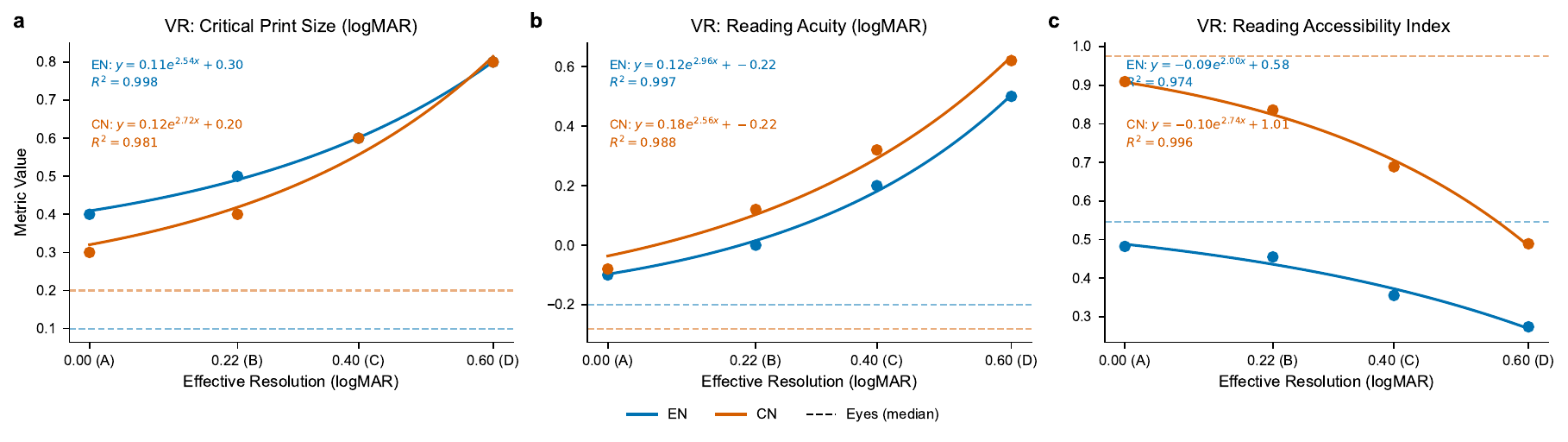}
\caption{Exponential fit of CPS, RA, and RAI (median) from VR condition as a function of effective resolution (logMAR). }
 \label{fig:resolution_mnread}
\end{figure*}

\begin{figure}[!ht]
 \centering\includegraphics[width=\columnwidth]{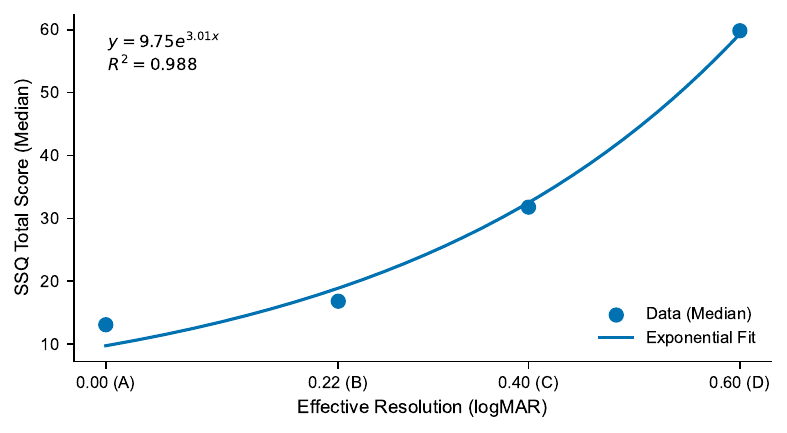}
\caption{Exponential fit of SSQ total score (median) from VR condition as a function of effective resolution (logMAR).}
 \label{fig:resolution_ssq}
\end{figure}

\subsection{Implementation of XR-Read}
The existing iPad implementation of MNREAD preserves the sentence design and logarithmic scaling while automating timing, scoring, and parameter estimation. Prior work has shown equivalence to the physical chart \cite{calabrese2018comparing}.
The sentence presentation and timing are initiated by a tap; the internal timer eliminates stopwatch variability and timestamps each trial. Errors are entered immediately after each sentence on a dedicated scoring pane. The device uses its known PPI together with the entered viewing distance to render requested angular sizes, so $S$ is computed directly rather than inferred from print metrics; the usable range is bounded by screen dimensions and resolution. At the standard 40 cm, the iPad’s PPI constrains measurement to $-0.1$ to $1.2$ logMAR (14 sentences in 0.1 logMAR steps). 

To measure reading acuity in both VR and VST, we developed 3D and 2D XR-Read in Unity. In both versions, the examiner advances to the next sentence by pressing the spacebar. For the font rendering, a signed distance field (SDF) font (TextMesh Pro) is used to provide crisp rendering under transformation and magnification. Times (EN) and SourceHan Serif Simplified Chinese (CN) were used to generate the SDF fonts according to the design of MNREAD (EN) and C-READ (CN). In 3D XR-Read for VR, the sentence panel is anchored to the eye camera and placed in 3D space 40 cm from the camera position (1 unit $=$ 100 cm in Unity) (see \figref{screenshots} (c) and (d)). In 2D XR-Read for VST, the sentence panel is rendered in 2D screen space and runs as a standalone Android application on a smartphone positioned 40 cm from the eyes (see \figref{photos} and \figref{screenshots} (a) and (b)). Print size is controlled by scaling the transform components. For the 2D XR-Read, we performed a physical calibration to obtain the device’s actual pixels/cm. XR-Read automatically records sentences, logMAR, and time, and outputs a CSV file. To avoid human error during the experiment, each session is audio-recorded; after the session, the examiner tallies errors from the recordings.

\subsection{Conditions and Procedure}
We conducted a within-subjects study manipulating three factors: language (CN, EN), display mode (VR, VST), and effective resolution (A $=$ 0.00, B $=$ 0.22, C $=$ 0.40, D $=$ 0.60 logMAR). Demographics and Pre-SSQ were collected at the beginning of the experiment. Participants were trained on the XR-Read of both languages as a trial before the formal study. The study began with a naked-eye condition using 2D XR-Read in both languages as the baseline. Participants then completed 16 tests spanning language $\times$ display mode $\times$ effective resolution, with order counterbalanced via Latin squares. For each display mode $\times$ effective resolution combination, a session comprised both languages and lasted about 2 minutes. Participants took sufficient breaks between sessions to prevent the accumulation of simulator sickness and completed the Post-SSQ after each session. The whole experiment is about 1 hour. In total, each language used 10 sets of 16 sentences (-0.5 to 1.0, at 0.1 logMAR steps): one trial set, one naked-eye set, and eight sets covering the display mode $\times$ effective resolution conditions. All sentences were unique and followed MNREAD and C-READ design principles \cite{han2017cread, mansfield2019extending}.
After completing all sessions and questionnaires, we conducted a brief semi-structured interview to capture participants' subjective experiences, with a focus on how they felt when the text became unreadable under VR and VST.

\subsection{Participants and Apparatus}
We recruited 16 participants (8 males, 8 females), ages 18--32 ($M=25.06$, $SD=3.53$). The experiments were classified as low-risk research, adhered to the university’s ethics guidelines and regulations, and received approval from the University Ethics Committee. All participants volunteered, provided informed consent, and had normal or corrected-to-normal vision. All were native Chinese speakers with professional academic English oral proficiency. Sessions were conducted in a closed laboratory at a constant temperature and normal luminance.

We used a high-end HMD, Varjo XR-4 Focal Edition with 55 PPD VR and 51 PPD VST, connected to a desktop with 32 GB RAM, a GeForce RTX 5080 GPU, and an Intel Core i9-14900KF CPU. The 2D XR-Read application ran on an Android smartphone, OnePlus 7T Pro with a 6.67-inch display, $1440 \times 3120$ pixels, 19.5:9 ratio, and 516 PPI (set to max brightness during the experiment). Its PPI is bigger than the iPad 3 (264 PPI) recommended by the MNREAD iPad app, enabling a -0.5 to 1.0 logMAR range (16 sentences at 0.1 logMAR steps) at 40 cm. A chinrest was used to maintain the 40 cm viewing distance.

\subsubsection{Calibration of Effective Resolution across VR and VST}
We calibrated the effective resolution for both VR and VST so that conditions were matched on a common logMAR scale. In VST, image quality is fixed and unaffected by render resolution scaling. Therefore, we introduced controlled optical blur by mounting trial lenses in front of the cameras and then measured physical visual acuity with a tumbling E chart. This procedure yielded four reference levels: 0.00 logMAR (no lens), 0.22 (0.2 for short) logMAR (-4.00 diopter (D)), 0.40 logMAR (-5.00 D), and 0.60 logMAR (-6.00 D). The 0.60 logMAR level approximates the median acuity reported for current consumer VST HMDs (i.e., Quest Pro, HTC Vive XR, Pico 4 Pro) according to research that used the same chart for VST HMDs \cite{ovva}.

For VR, we varied the render resolution scale in Unity XR (0.10, 0.40, 0.70, 1.00) and measured central visual acuity with the OVVA test. A logarithmic regression described the relationship between scale x and acuity y (logMAR): \(y=-0.2796\ln(x)-0.0232\). Using this function, we identified the scales corresponding to the four target acuity levels, mapping \(y\in\{0.00, 0.22, 0.40, 0.60\}\) to \(x\approx\{0.92, 0.42, 0.22, 0.11\}\) (see \figref{resolution_curve}).

In summary, VST conditions were produced via optical blur and VR conditions via render scaling; both were expressed in the same set of logMAR values, enabling direct comparison across display modes.

\subsection{Results}
Across analyses, effective resolution strongly modulated reading performance and symptoms, with larger and more consistent effects in VR than in VST; language (CN vs. EN) and display mode (VR vs. VST) each contributed additional, resolution-dependent differences. Because distributions were non-normal, all data were non-parametric. For factors with $k\geq 3$ levels (e.g., \emph{Effective Resolution} A--D), we used Friedman tests followed by Wilcoxon signed-rank post hoc comparisons with Bonferroni correction; for $k=2$ factors (e.g., \emph{Display} and \emph{Language}: CN vs. EN), we used Wilcoxon signed-rank tests. Where possible, we report effect sizes, using $W$ for Friedman tests to index magnitude.

\subsubsection{Effects of Effective Resolution on Reading Performance}
For reading performance across effective resolution, resolution changes produced robust performance differences, especially in VR. Friedman tests indicated that nearly all metrics (ACC, CPS, MRS, and RA) in VR varied significantly with resolution (EN MRS was the lone exception), with medium to large effect sizes; in VST, effects were weaker but still significant for all metrics with small to medium effect sizes (see \tableref{MNREAD_Friedman}). 

Post hoc tests clarified a monotonic trend: in VR, ACC decreased while CPS and RA increased as resolution worsened (A$\rightarrow$D); VST showed the same qualitative pattern but with fewer significant pairwise differences, most prominently separating the most degraded level (D) from higher-quality levels (A--C) (see \tableref{MNREAD_Posthoc_resolution_CN}, \tableref{MNREAD_Posthoc_resolution_EN}, and \figref{MNREAD}). Relative to median naked eye performance, MRS was most similar, whereas other metrics were generally worse at most resolutions and approached the eye only at the best level (A) (\figref{MNREAD}).

\subsubsection{Display Mode Comparisons and Language Effects}
For display mode differences (VR vs. VST), at matched logMAR, VST underperformed VR on several outcomes. Wilcoxon contrasts showed significantly poorer performance in VST for CPS, RA, and ACC at multiple resolutions, with a CN-specific decrement in MRS (\tableref{XRRead_combined_horizontal_simplified_updated}, right block).

For language differences (CN vs. EN), across all resolutions, CN yielded better reading outcomes than EN on key metrics. Wilcoxon tests revealed significantly higher ACC and better (lower) MRS for CN at every resolution (all \(p<.001\)) (\tableref{XRRead_combined_horizontal_simplified_updated}, left block).

\subsubsection{Simulator Sickness Questionnaire Outcomes}
Resolution degradations increased SSQ scores, particularly in VR. Friedman tests showed significant effects of resolution in VR for Nausea, Oculomotor, Disorientation, and Total Score, and in VST for Nausea, Oculomotor, and Total Score (but not Disorientation) (\tableref{SSQ_Friedman}). Post hoc comparisons revealed numerous significant pairwise increases with worsening resolution, with most differences emerging outside the A vs. B contrast (\tableref{SSQ_Posthoc_resolution}). 

\subsubsection{Exponential Modeling of Effective Resolution}
Exponential models captured resolution-dependence in VR with excellent fits, indicating rapid improvement toward eye performance as effective resolution improved. Specifically, CPS increased with logMAR (EN: \(y=0.1091\,e^{2.5400x}+0.3000,\,R^2=.9981\); CN: \(y=0.1202\,e^{2.7200x}+0.2000,\,R^2=.9807\)), RA increased exponentially (EN: \(y=0.1228\,e^{2.9600x}-0.2200,\,R^2=.9970\); CN: \(y=0.1838\,e^{2.5600x}-0.2200,\,R^2=.9880\)), and ACC decreased exponentially (EN: \(y=-0.0942\,e^{2.0000x}+0.5822,\,R^2=.9739\); CN: \(y=-0.1015\,e^{2.7400x}+1.0093,\,R^2=.9956\)). While CPS, RA, and ACC asymptotically approach eye performance at better resolutions, even level A remained measurably worse than the naked eye (\figref{resolution_mnread}).

An exponential model for VR SSQ Total Score confirmed rapid growth with degradation (\(y=9.7505\,e^{3.0099x},\,R^2=.988\)) and indicated that even the best resolution (A) produced non-zero simulator sickness symptoms (\figref{resolution_ssq}). Direct VR vs. VST comparisons showed consistently higher Oculomotor, Disorientation, and Total Score in VST at multiple resolutions (\tableref{SSQ_Posthoc_display}).

\subsubsection{Qualitative Findings}
We conducted brief semi-structured interviews at the end of the study to capture subjective experiences in both VR and VST when text became unreadable. Overall, participants reported a consistent set of perceptual, physical, and emotional responses that intensified as effective resolution degraded from A to D. Participants frequently described \emph{flicker}, \emph{text shimmering}, \emph{jitter/oscillation}, \emph{difficulty focusing}, and occasional \emph{double images/ghosting}. Several likened severe blur to ``mosaic'' text. As blur increased, letters were perceived as \emph{jumping}, \emph{shrinking}, or \emph{disappearing into the distance}. Many noted an impression of \emph{pixelation} or insufficient pixel density (more salient in VR). Moreover, participants reported \emph{eyestrain}, \emph{dryness from reduced blinking}, \emph{headache}, and \emph{dizziness}; some mentioned \emph{nausea} at the worst levels. Several described \emph{squinting} or wanting to \emph{rub the eyes}.

VR was \emph{slightly better} overall than VST for many participants since \emph{small phone screen and difficulty focusing} under VST made fatigue and effort more salient and sometimes ``more tiring’’ despite fewer rendering artifacts. Symptoms and negative affect increased with worse effective resolution. The interviews qualitatively align with SSQ patterns: as effective resolution worsens, oculomotor strain, disorientation/dizziness, and nausea increase, with artifacts (flicker, pixelation, shimmer) more salient in VR and fatigue from small display size and difficulty focusing more salient in VST. Participants’ coping behaviors (squinting, reduced blinking) may contribute to the observed SSQ increases.

\section{Discussion}
Our study set out to quantify how end-to-end \emph{effective resolution} (expressed in logMAR) affects functional reading performance and comfort in XR. Here, we interpret the results with respect to the three research questions.

\subsection{RQ1: How does effective resolution modulate reading performance?}
Across both languages and display modes, degrading effective resolution from $0.0$ to $0.6$~~logMAR produced substantial changes in all reading outcomes (\figref{MNREAD}, \tableref{MNREAD_Friedman},\tableref{MNREAD_Posthoc_resolution_CN},\tableref{MNREAD_Posthoc_resolution_EN}). In VR, these effects were large and highly reliable: CPS and RA rose steeply with logMAR while ACC fell, with exponential models providing excellent fits (\figref{resolution_mnread}). Notably, EN MRS in VR did not reach significance, suggesting that reading rate is comparatively robust until clarity crosses a threshold, after which CPS/RA become the limiting factors.

Two patterns are noteworthy. First, even at the best effective resolution condition (A: $0$ logMAR), performance had not fully converged to naked-eye baselines (\figref{resolution_mnread}). This residual gap likely reflects XR-specific factors that persist at high nominal clarity (e.g., optics, temporal artifacts, binocular alignment/vergence demand, font rasterization differences). Second, performance changes with resolution were smooth and predictable for VR. The exponential fits provide reference curves that map a target logMAR to expected CPS/RA/ACC ranges, enabling forward design of type scales and rendering budgets.

\subsection{RQ2: Do these effects generalize across display modes and languages?}
Our results confirm that these effects generalize across display modes and languages. For display mode, when matched by logMAR, VST still underperformed VR for several metrics (CPS, RA, ACC; and MRS for CN at multiple levels; \tableref{XRRead_combined_horizontal_simplified_updated}). Resolution, therefore, explains most but not all of the variance: the VST pipeline adds penalties (camera optics, sensor noise/sharpening, tone mapping, view stabilization, capture–display latency) that are not captured by acuity alone. Practically, this means that meeting a given logMAR target in VST is necessary but may be insufficient for VR-equivalent readability and comfort; designers need to account for additional performance loss in VST beyond the acuity level.

For language, the direction of effects generalized across English and Chinese, but absolute levels differed. CN yielded higher MRS and ACC at all resolutions (all $p<.001$), with similar resolution-dependent slopes. Because participants were native CN speakers with professional EN proficiency, we interpret the language differences primarily as reader proficiency effects. The key result for generalization is the shared functional form: regardless of language, worse logMAR reliably raised CPS/RA and reduced ACC.

\subsection{RQ3: How does effective resolution relate to simulator sickness, and where are practical operating points?}
SSQ scores rose sharply as logMAR worsened, especially in VR, where Total Score increased exponentially with excellent fit (\figref{resolution_ssq},\tableref{SSQ_Friedman},\tableref{SSQ_Posthoc_resolution}). Post-hoc tests showed many significant pairwise increases between adjacent resolution levels, suggesting that comfort declines quickly as resolution decreases. VST produced higher Oculomotor and Disorientation and Total Score than VR at multiple resolutions (\tableref{SSQ_Posthoc_display}), consistent with added camera/display processing overhead.

Combining performance and SSQ trends reveals a clarity tipping point near \textbf{0 logMAR} \textit{(normal visual acuity)}. At this level of clarity or better, reading performance approaches naked eye, and simulator sickness grows slowly; at poorer clarity, bigger losses incur steep costs in both performance and comfort. Even the best effective resolution (0 logMAR) produced non-zero simulator sickness, underscoring that clarity is a major driver but not the sole determinant of comfort. The qualitative reports (flicker, shimmer, oscillation, effort) align with oculomotor and disorientation increases, suggesting that defects and accommodation/vergence stress compound the effects of reduced spatial fidelity.

We consolidate all actionable recommendations based on these findings in the following section.

\section{Design and evaluation implications} \label{sec:design}
Based on the findings, we propose four key design principles for XR HMDs and their applications. 

\begin{enumerate}
\item \textbf{Clarity baseline and targets.} Treat effective resolution (logMAR) as the primary control knob and enforce \textbf{0 logMAR} (normal visual acuity) as a baseline requirement for sustained, productivity-grade reading. At this level of clarity or better, reading performance approaches naked eye, and simulator sickness growth is modest; poorer clarity produces steep costs in both performance and comfort. Designers can use the fitted reference curves to map target logMAR to expected read performance when setting type scales and render budgets.

\item \textbf{Mode-aware budgeting (VST vs. VR).} Matching by logMAR is necessary but not sufficient for VST. The VST pipeline (camera optics, sensor noise/sharpening, tone mapping, capture–display latency) imposes penalties not captured by acuity alone. To achieve VR-comparable readability and comfort in VST, \textbf{budget extra margin and target $\leq 0$ logMAR or better}.

\item \textbf{Cross-mode Quality Assurance (QA).} Validate readability and simulator sickness under both VR and VST at matched logMAR; do not assume equivalence. Incorporate XR-Read runs into regression testing and track reading performance (CPS, RA, ACC, MRS) alongside simulator sickness. Use matched-logMAR test points to isolate non-acuity penalties in VST.

\item \textbf{Comfort budgets.} Readability does not guarantee comfort. Treat simulator sickness growth with logMAR as a cost function when planning session length, layout density, and motion/animation rates. Establish acceptance bands for simulator sickness scales at the intended operating logMAR and defects (flicker/shimmer/oscillation) during QA.

\end{enumerate}

\section{Limitations and Future Work}
Our findings should be interpreted with several limitations in mind. Participants (N=16) were young adults (18–32), all native CN speakers with professional EN proficiency. Language effects likely reflect proficiency; generalization to other demographics, age groups, and clinical populations remains to be tested. 

Our results were obtained with one high-end HMD (Varjo XR-4 Focal Edition) and a single phone for 2D XR-Read under VST. Although we calibrated both modes to match logMAR, the effects of device-specific MTF and rendering strategies remains to be tested. Replication across HMDs, optics, and rendering pipelines is a priority. VST clarity was controlled via external trial lenses in front of cameras. While this produced calibrated acuity levels, the induced blur differs from native VST limitations (e.g., sensor noise, demosaicing, rolling shutter). Future work should combine optical and electronic degradations and include direct-view passthrough pipelines. XR-Read uses MNREAD/C-READ sentences and timed reading. Real-world productivity involves diverse text (browser, code, spreadsheets, long-form articles) and interactions (scrolling, head/eye movements). Extending to dynamic layouts, varied fonts/scripts, contrast and luminance manipulations, and longer sessions will improve validity. We also plan to broaden language/script coverage and proficiency levels and embed XR-Read into XR apps/hardware in the future.

\section{Conclusion}
In this work, we introduced \emph{XR-Read}, a language-switchable reading acuity test that calibrates and manipulates \emph{effective} resolution in VR and VST on a unified logMAR scale. Using a within-subjects study ($N=16$), we showed that reducing effective resolution systematically slowed reading (MRS), enlarged critical print size and acuity thresholds (CPS, RA), lowered accessibility (ACC), and increased simulator sickness. These effects followed consistent, often exponential, patterns across display modes and languages, with VST carrying additional penalties and language proficiency shaping absolute performance. 

Together with generalized reference curves, our results highlight \textit{normal visual acuity} ($0$ logMAR) as a practical baseline: at this level of clarity or better, reading performance approaches naked-eye levels and simulator sickness remains modest, whereas poorer clarity rapidly degrades both. We therefore elevate effective resolution as a key system-level lever for XR usability, suggesting $\leq 0$ logMAR as a design requirement for XR systems intended for daily, text-heavy use. Designers should allocate headroom for VST-specific penalties and validate against this threshold with repeatable, language-aware tests. These findings establish actionable targets for making XR a viable platform for everyday reading and productivity.

\addtocontents{toc}{\protect\setcounter{tocdepth}{-1}}

\bibliographystyle{unsrtnat}

\putbib[MyCollection]

\section*{Author contributions statement}

J.W. proposed the project, J.W., X.C and B.H. conceived the experiment. J.W., X.C and B.H. designed the psychophysical study and conducted the experiment under H.L.'s supervision. J.W. analysed the results and created the figures with feedback from all authors. All authors contributed to writing and reviewing the manuscript.

\end{bibunit}

\end{document}

%% file: macros.tex
\usepackage{url}
\usepackage{xspace}
\usepackage{comment}
\usepackage{amsmath}

\newcommand{\figref}[1]{Figure~\ref{fig:#1}}

\newcommand{\tableref}[1]{Table~\ref{tab:#1}}

\renewcommand{\eqref}[1]{Eq.~\ref{eq:#1}}

\usepackage{placeins}

\usepackage{subcaption} 

\usepackage{multirow}
\usepackage[table]{xcolor}






%% file: main.bbl
\begin{thebibliography}{43}
\providecommand{\natexlab}[1]{#1}
\providecommand{\url}[1]{\texttt{#1}}
\expandafter\ifx\csname urlstyle\endcsname\relax
  \providecommand{\doi}[1]{doi: #1}\else
  \providecommand{\doi}{doi: \begingroup \urlstyle{rm}\Url}\fi

\bibitem[Vasarainen et~al.(2021)Vasarainen, Paavola, and Vetoshkina]{vasarainen2021systematic}
Minna Vasarainen, Sami Paavola, and Liubov Vetoshkina.
\newblock {A systematic literature review on extended reality: virtual, augmented and mixed reality in working life}.
\newblock \emph{Internationa Journal of Virtual Reality}, 21\penalty0 (2):\penalty0 1--28, 2021.

\bibitem[Domingues et~al.(2024)Domingues, Vieira, Yoshida, de~Oliveira, Peres, Mauricio, Nunes, Teixeira, and Neto]{10.1145/3691573.3691599}
Gustavo~Camargo Domingues, Vitor Vieira, Leina Yoshida, Leticia de~Oliveira, Fabiana Peres, Claudio Mauricio, Fatima Nunes, Jo{\~{a}}o~Marcelo Teixeira, and Amadeo Neto.
\newblock {What if Video See-Through in HMDs Changes How Accurately We Perform Tasks?}
\newblock In \emph{Proceedings of the 26th Symposium on Virtual and Augmented Reality}, SVR '24, pages 213--222, New York, NY, USA, 2024. Association for Computing Machinery.
\newblock ISBN 9798400709791.
\newblock \doi{10.1145/3691573.3691599}.
\newblock URL \url{https://doi.org/10.1145/3691573.3691599}.

\bibitem[Grout et~al.(2015)Grout, Rogers, Apperley, and Jones]{grout2015reading}
Cameron Grout, William Rogers, Mark Apperley, and Steve Jones.
\newblock {Reading text in an immersive head-mounted display: An investigation into displaying desktop interfaces in a 3D virtual environment}.
\newblock In \emph{Proceedings of the 15th New Zealand conference on human-computer interaction}, pages 9--16, 2015.

\bibitem[Knierim et~al.(2020)Knierim, Kosch, Groschopp, and Schmidt]{10.1145/3334480.3382920}
Pascal Knierim, Thomas Kosch, Johannes Groschopp, and Albrecht Schmidt.
\newblock {Opportunities and Challenges of Text Input in Portable Virtual Reality}.
\newblock In \emph{Extended Abstracts of the 2020 CHI Conference on Human Factors in Computing Systems}, CHI EA '20, pages 1--8, New York, NY, USA, 2020. Association for Computing Machinery.
\newblock ISBN 9781450368193.
\newblock \doi{10.1145/3334480.3382920}.
\newblock URL \url{https://doi.org/10.1145/3334480.3382920}.

\bibitem[Koji{\'{c}} et~al.(2020)Koji{\'{c}}, Ali, Greinacher, M{\"{o}}ller, and Voigt-Antons]{9123091}
Tanja Koji{\'{c}}, Danish Ali, Robert Greinacher, Sebastian M{\"{o}}ller, and Jan-Niklas Voigt-Antons.
\newblock {User Experience of Reading in Virtual Reality — Finding Values for Text Distance, Size and Contrast}.
\newblock In \emph{2020 Twelfth International Conference on Quality of Multimedia Experience (QoMEX)}, pages 1--6, 2020.
\newblock \doi{10.1109/QoMEX48832.2020.9123091}.

\bibitem[de~Souza and Tartz(2024)]{desouza2024vsteval}
Jessica de~Souza and Robert Tartz.
\newblock Visual perception and user satisfaction in video see-through head-mounted displays: a mixed-methods evaluation.
\newblock \emph{Frontiers in Virtual Reality}, 5:\penalty0 1368721, 2024.
\newblock \doi{10.3389/frvir.2024.1368721}.

\bibitem[Stauffert et~al.(2020)Stauffert, Niebling, and Latoschik]{Stauffert2020Latency}
Jan-Philipp Stauffert, Florian Niebling, and Marc~Erich Latoschik.
\newblock Latency and cybersickness: Impact, causes, and measures. a review.
\newblock \emph{Frontiers in Virtual Reality}, 1:\penalty0 582204, 2020.
\newblock \doi{10.3389/frvir.2020.582204}.

\bibitem[Palmisano et~al.(2020)Palmisano, Mursic, and Kim]{Palmisano2020DVP}
Stephen Palmisano, Renata Mursic, and Juno Kim.
\newblock Cybersickness in head-mounted displays is caused by differences in the user's virtual and physical head pose.
\newblock \emph{Frontiers in Virtual Reality}, 1:\penalty0 587698, 2020.
\newblock \doi{10.3389/frvir.2020.587698}.

\bibitem[Hoffman et~al.(2008)Hoffman, Girshick, Akeley, and Banks]{Hoffman2008VAC}
David~M. Hoffman, Ahna~R. Girshick, Kurt Akeley, and Martin~S. Banks.
\newblock Vergence--accommodation conflicts hinder visual performance and cause visual fatigue.
\newblock \emph{Journal of Vision}, 8\penalty0 (3):\penalty0 33, 2008.
\newblock \doi{10.1167/8.3.33}.

\bibitem[Shibata et~al.(2011)Shibata, Kim, Hoffman, and Banks]{Shibata2011Zone}
Takashi Shibata, Joohwan Kim, David~M. Hoffman, and Martin~S. Banks.
\newblock The zone of comfort: Predicting visual discomfort with stereo displays.
\newblock \emph{Journal of Vision}, 11\penalty0 (8):\penalty0 11, 2011.
\newblock \doi{10.1167/11.8.11}.

\bibitem[Souchet et~al.(2022)Souchet, Philippe, Lourdeaux, and Leroy]{souchet2022measuring}
Alexis~D Souchet, St{\'{e}}phanie Philippe, Domitile Lourdeaux, and Laure Leroy.
\newblock {Measuring visual fatigue and cognitive load via eye tracking while learning with virtual reality head-mounted displays: A review}.
\newblock \emph{International Journal of Human--Computer Interaction}, 38\penalty0 (9):\penalty0 801--824, 2022.

\bibitem[Erickson et~al.(2020)Erickson, Kim, Bruder, and Welch]{Erickson2020}
Austin Erickson, Kangsoo Kim, Gerd Bruder, and Gregory~F. Welch.
\newblock Effects of dark mode graphics on visual acuity and fatigue with virtual reality head-mounted displays.
\newblock In \emph{2020 IEEE Conference on Virtual Reality and 3D User Interfaces (VR)}, pages 434--442, 2020.
\newblock \doi{10.1109/VR46266.2020.00064}.

\bibitem[Lu et~al.(2023)Lu, Lian, and Jia]{lu2023display}
Jiawei Lu, Trisha Lian, and Jerry Jia.
\newblock {Display and imaging system sharpness modeling and requirement in high-resolution VR and AR}.
\newblock \emph{Electronic Imaging}, 35\penalty0 (12):\penalty0 211--213, 2023.

\bibitem[Wang et~al.(2024)Wang, Shi, Li, Wei, and Liang]{ovva}
Jialin Wang, Rongkai Shi, Xiaodong Li, Yushi Wei, and Hai-Ning Liang.
\newblock {Omnidirectional Virtual Visual Acuity: A User-Centric Visual Clarity Metric for Virtual Reality Head-Mounted Displays and Environments}.
\newblock \emph{IEEE Transactions on Visualization and Computer Graphics}, 30\penalty0 (5):\penalty0 2033--2043, 2024.
\newblock \doi{10.1109/TVCG.2024.3372127}.

\bibitem[Ashraf et~al.(2025)Ashraf, Chapiro, and Mantiuk]{ashraf2025resolution}
Maliha Ashraf, Alexandre Chapiro, and Rafa{\l}~K Mantiuk.
\newblock Resolution limit of the eye—how many pixels can we see?
\newblock \emph{Nature Communications}, 16\penalty0 (1):\penalty0 9086, 2025.

\bibitem[Monk(2014)]{monk2014fundamentals}
Andrew~F Monk.
\newblock \emph{{Fundamentals of human-computer interaction}}.
\newblock Academic Press, 2014.

\bibitem[Clinton(2019)]{Clinton2019}
Virginia Clinton.
\newblock {Reading from paper compared to screens: A systematic review and meta-analysis}.
\newblock \emph{Journal of Research in Reading}, 42\penalty0 (2):\penalty0 288--325, 2019.
\newblock ISSN 14679817.
\newblock \doi{10.1111/1467-9817.12269}.

\bibitem[Legge(2006)]{legge2006psychophysics}
Gordon~E Legge.
\newblock \emph{{Psychophysics of reading in normal and low vision}}.
\newblock CRC Press, 2006.

\bibitem[Legge et~al.(1989)Legge, Ross, Luebker, and LaMay]{legge1989MNREAD}
Gordon~E. Legge, Julie~A. Ross, Angela Luebker, and John~M. LaMay.
\newblock Psychophysics of reading. viii. the minnesota low-vision reading test.
\newblock \emph{Optometry and Vision Science}, 66\penalty0 (12):\penalty0 843--853, 1989.

\bibitem[Mansfield et~al.(1993)Mansfield, Ahn, Legge, and Luebker]{Mansfield1993MNREAD}
J.~Stephen Mansfield, Sonia~J. Ahn, Gordon~E. Legge, and Andrew Luebker.
\newblock A new reading-acuity chart for normal and low vision.
\newblock In \emph{Noninvasive Assessment of the Visual System (NAVS 1993)}, Optics InfoBase Conference Papers, pages 232--235. Optica Publishing Group, 1993.

\bibitem[Rubin(2013)]{Rubin2013Measuring}
Gary~S. Rubin.
\newblock Measuring reading performance.
\newblock \emph{Vision Research}, 90:\penalty0 43--51, 2013.
\newblock \doi{10.1016/j.visres.2013.02.015}.

\bibitem[Calabr\`{e}se et~al.(2016)Calabr\`{e}se, Owsley, McGwin, and Legge]{calabrese2016ACC}
Aur\'{e}lie Calabr\`{e}se, Cynthia Owsley, Gerald McGwin, and Gordon~E. Legge.
\newblock Development of a reading accessibility index using the mnread acuity chart.
\newblock \emph{JAMA Ophthalmology}, 134\penalty0 (4):\penalty0 398--405, 2016.
\newblock \doi{10.1001/jamaophthalmol.2015.6097}.

\bibitem[Weir et~al.(2023)Weir, Loizides, Nahar, Aggoun, and Pollard]{weir2023see}
Kurtis Weir, Fernando Loizides, Vinita Nahar, Amar Aggoun, and Andrew Pollard.
\newblock {I see therefore i read: improving the reading capabilities of individuals with visual disabilities through immersive virtual reality}.
\newblock \emph{Universal Access in the Information Society}, 22\penalty0 (2):\penalty0 387--413, 2023.

\bibitem[Han et~al.(2017)Han, Cong, Yu, and Liu]{han2017cread}
Qi-Ming Han, Lin-Juan Cong, Cong Yu, and Lei Liu.
\newblock Developing a logarithmic chinese reading acuity chart.
\newblock \emph{Optometry and Vision Science}, 94\penalty0 (6):\penalty0 714--724, 2017.
\newblock \doi{10.1097/OPX.0000000000001081}.

\bibitem[Calabrese et~al.(2018)Calabrese, To, He, Berkholtz, Rafian, and Legge]{calabrese2018comparing}
Aurelie Calabrese, Long To, Yingchen He, Elizabeth Berkholtz, Paymon Rafian, and Gordon~E Legge.
\newblock {Comparing performance on the MNREAD iPad application with the MNREAD acuity chart}.
\newblock \emph{Journal of vision}, 18\penalty0 (1):\penalty0 8, 2018.

\bibitem[Chaurasia et~al.(2020)Chaurasia, Nieuwoudt, Ichim, Szeliski, and Sorkine-Hornung]{PassthroughPlus}
Gaurav Chaurasia, Arthur Nieuwoudt, Alexandru-Eugen Ichim, Richard Szeliski, and Alexander Sorkine-Hornung.
\newblock {Passthrough+: Real-time Stereoscopic View Synthesis for Mobile Mixed Reality}.
\newblock \emph{Proc. ACM Comput. Graph. Interact. Tech.}, 3\penalty0 (1), 2020.
\newblock \doi{10.1145/3384540}.
\newblock URL \url{https://doi.org/10.1145/3384540}.

\bibitem[Pohl et~al.(2013)Pohl, Johnson, and Bolkart]{Pohl2013}
Daniel Pohl, Gregory~S. Johnson, and Timo Bolkart.
\newblock {Improved pre-warping for wide angle, head mounted displays}.
\newblock \emph{Proceedings of the ACM Symposium on Virtual Reality Software and Technology, VRST}, 2\penalty0 (2):\penalty0 259--262, 2013.
\newblock \doi{10.1145/2503713.2503752}.

\bibitem[Carkeet et~al.(2021)Carkeet, Lister, and Goh]{Carkeet2021}
Andrew Carkeet, Lucas Lister, and Yee~Teng Goh.
\newblock {Computer monitor pixellation and Landolt C visual acuity}.
\newblock \emph{Ophthalmic and Physiological Optics}, 41\penalty0 (6):\penalty0 1176--1182, 2021.
\newblock ISSN 14751313.
\newblock \doi{10.1111/opo.12882}.

\bibitem[Xiao et~al.(2022)Xiao, Nouri, Hegland, Garcia, and Lanman]{NeuralPassthrough}
Lei Xiao, Salah Nouri, Joel Hegland, Alberto~Garcia Garcia, and Douglas Lanman.
\newblock {NeuralPassthrough: Learned Real-Time View Synthesis for VR}.
\newblock In \emph{ACM SIGGRAPH 2022 Conference Proceedings}, SIGGRAPH '22, New York, NY, USA, 2022. Association for Computing Machinery.
\newblock ISBN 9781450393379.
\newblock \doi{10.1145/3528233.3530701}.
\newblock URL \url{https://doi.org/10.1145/3528233.3530701}.

\bibitem[Cheung et~al.(2015)Cheung, Liu, Lam, and Cheong]{cheung2015chinesereading}
Josephine P.~Y. Cheung, Dilys S.~K. Liu, Catherine C.~C. Lam, and Allen M.~Y. Cheong.
\newblock Development and validation of a new chinese reading chart for children.
\newblock \emph{Ophthalmic and Physiological Optics}, 35\penalty0 (5):\penalty0 514--521, 2015.
\newblock \doi{10.1111/opo.12228}.

\bibitem[Mansfield et~al.(2019)Mansfield, Atilgan, Lewis, and Legge]{mansfield2019extending}
J~Stephen Mansfield, Nilsu Atilgan, A~M Lewis, and G~E Legge.
\newblock {Extending the MNREAD sentence corpus: Computer-generated sentences for measuring visual performance in reading}.
\newblock \emph{Vision research}, 158:\penalty0 11--18, 2019.

\bibitem[Kilpel\"ainen and H\"akkinen(2023)]{kilpelainen2023xrlegibility}
Markku Kilpel\"ainen and Jukka H\"akkinen.
\newblock An effective method for measuring text legibility in xr devices reveals clear differences between three devices.
\newblock \emph{Frontiers in Virtual Reality}, 4:\penalty0 1243387, 2023.
\newblock \doi{10.3389/frvir.2023.1243387}.

\bibitem[Novotny and Laidlaw(2024)]{novotny2024readingvr}
Johannes Novotny and David~H. Laidlaw.
\newblock Evaluating text reading speed in vr scenes and 3d particle visualizations.
\newblock \emph{IEEE Transactions on Visualization and Computer Graphics}, 30\penalty0 (5):\penalty0 2602--2612, 2024.
\newblock \doi{10.1109/TVCG.2024.3372093}.

\bibitem[Vona et~al.(2025)Vona, Schorlemmer, Stern, Ashrafi, Vergari, Kojic, and Voigt-Antons]{10972690}
Francesco Vona, Julia Schorlemmer, Michael Stern, Navid Ashrafi, Maurizio Vergari, Tanja Kojic, and Jan-Niklas Voigt-Antons.
\newblock {Comparing Pass-Through Quality of Mixed Reality Devices: A User Experience Study During Real-World Tasks}.
\newblock In \emph{2025 IEEE Conference on Virtual Reality and 3D User Interfaces Abstracts and Workshops (VRW)}, pages 1432--1433, 2025.
\newblock \doi{10.1109/VRW66409.2025.00362}.

\bibitem[Kennedy et~al.(1993)Kennedy, Lane, Berbaum, and Lilienthal]{kennedy1993ssq}
Robert~S. Kennedy, Norman~E. Lane, Kevin~S. Berbaum, and Michael~G. Lilienthal.
\newblock Simulator sickness questionnaire: An enhanced method for quantifying simulator sickness.
\newblock \emph{The International Journal of Aviation Psychology}, 3\penalty0 (3):\penalty0 203--220, 1993.

\bibitem[Saredakis et~al.(2020)Saredakis, Szpak, Birckhead, Keage, Rizzo, and Loetscher]{saredakis2020meta}
Dimitrios Saredakis, Ancret Szpak, Brandon Birckhead, Hannah A.~D. Keage, Albert Rizzo, and Tobias Loetscher.
\newblock Factors associated with virtual reality sickness in head-mounted displays: A systematic review and meta-analysis.
\newblock \emph{Frontiers in Human Neuroscience}, 14:\penalty0 96, 2020.
\newblock \doi{10.3389/fnhum.2020.00096}.

\bibitem[Fernandes and Feiner(2016)]{Fernandes2016}
Ajoy~S. Fernandes and Steven~K. Feiner.
\newblock {Combating VR sickness through subtle dynamic field-of-view modification}.
\newblock \emph{2016 IEEE Symposium on 3D User Interfaces, 3DUI 2016 - Proceedings}, pages 201--210, 2016.
\newblock \doi{10.1109/3DUI.2016.7460053}.

\bibitem[Wang et~al.(2023)Wang, Shi, Zheng, Xie, Kao, and Liang]{wang2023effect}
Jialin Wang, Rongkai Shi, Wenxuan Zheng, Weijie Xie, Dominic Kao, and Hai-Ning Liang.
\newblock {Effect of Frame Rate on User Experience, Performance, and Simulator Sickness in Virtual Reality}.
\newblock \emph{IEEE Transactions on Visualization and Computer Graphics}, 29\penalty0 (5):\penalty0 2478--2488, 2023.

\bibitem[Holladay(2004)]{holladay2004visual}
Jack~T Holladay.
\newblock {Visual acuity measurements}.
\newblock \emph{Journal of Cataract \& Refractive Surgery}, 30\penalty0 (2):\penalty0 287--290, 2004.

\bibitem[Cheng et~al.(2023)Cheng, Yao, Xu, Dai, Qin, Ye, Suo, and Zhang]{cheng2023using}
Tian Cheng, Taikang Yao, Boxuan Xu, Wanwei Dai, Xuejiao Qin, Juan Ye, Lingge Suo, and Chun Zhang.
\newblock {Using the C-Read as a Portable Device to Evaluate Reading Ability in Young Chinese Adults: An Observational Study}.
\newblock \emph{Journal of Personalized Medicine}, 13\penalty0 (3):\penalty0 463, 2023.

\bibitem[Baskaran et~al.(2019)Baskaran, Macedo, He, Hernandez-Moreno, Queir{\'o}s, Mansfield, and Calabr{\`e}se]{baskaran2019scoring}
Karthikeyan Baskaran, Antonio~Filipe Macedo, Yingchen He, Laura Hernandez-Moreno, Tatiana Queir{\'o}s, J~Stephen Mansfield, and Aur{\'e}lie Calabr{\`e}se.
\newblock Scoring reading parameters: an inter-rater reliability study using the mnread chart.
\newblock \emph{PloS one}, 14\penalty0 (6):\penalty0 e0216775, 2019.

\bibitem[Xiong et~al.(2018)Xiong, Lorsung, Mansfield, Bigelow, and Legge]{xiong2018reading2}
Ying-Zi Xiong, Ethan~A Lorsung, John~Stephen Mansfield, Charles Bigelow, and Gordon~E Legge.
\newblock Fonts designed for macular degeneration: Impact on reading.
\newblock \emph{Investigative ophthalmology \& visual science}, 59\penalty0 (10):\penalty0 4182--4189, 2018.

\bibitem[Calabrese et~al.(2016)Calabrese, Cheong, Cheung, He, Kwon, Mansfield, Subramanian, Yu, and Legge]{calabrese2016baseline}
Aur{\'{e}}lie Calabrese, Allen M~Y Cheong, Sing-Hang Cheung, Yingchen He, MiYoung Kwon, J~Stephen Mansfield, Ahalya Subramanian, Deyue Yu, and Gordon~E Legge.
\newblock {Baseline MNREAD measures for normally sighted subjects from childhood to old age}.
\newblock \emph{Investigative ophthalmology \& visual science}, 57\penalty0 (8):\penalty0 3836--3843, 2016.

\end{thebibliography}
